\documentclass[12pt]{article}
\usepackage{epsfig}
\usepackage{amssymb}

\def\b{\beta}

\def\d{\delta}

\def\k{\kappa}

\def\n{\nu}
\def\o{\omega}

\def\r{\rho}

\def\D{\Delta}

\def\G{\Gamma}

\def\O{\Omega}

\def\ve{\varepsilon}

\def\mt{\widetilde{m}_1}

\def\mtt{\widetilde{m}_2}

\def\be{\begin{equation}}
\def\ee{\end{equation}}
\def\ds{\displaystyle}

\def\bea{\begin{eqnarray}}
\def\eea{\end{eqnarray}}

\def\pl#1#2#3{Phys.~Lett.~{\bf B {#1}} ({#2}) #3}
\def\np#1#2#3{Nucl.~Phys.~{\bf B {#1}} ({#2}) #3}
\def\prl#1#2#3{Phys.~Rev.~Lett.~{\bf #1} ({#2}) #3}
\def\pr#1#2#3{Phys.~Rev.~{\bf D {#1}} ({#2}) #3}

\begin{document}
\date{}

\title{
{\normalsize
\mbox{ }\hfill
\begin{minipage}{3cm}
MPP-2005-5
\end{minipage}}\\
\vspace{5mm}
\bf Seesaw geometry and leptogenesis}
\author{P. Di Bari \\
{\it Max-Planck-Institut f\"{u}r Physik} \\ {\it (Werner-Heisenberg-Institut)} \\
{\it F\"{o}hringer Ring 6, 80805 M\"{u}nchen}}

\maketitle

\thispagestyle{empty}
\centerline{\date{\today}}

\begin{abstract}
\noindent
The representation of the seesaw orthogonal matrix in
the complex plane establishes a graphical correspondence between
neutrino mass models  and geometrical configurations, particularly
useful to study relevant aspects of leptogenesis. We first derive
the $C\!P$ asymmetry bound for hierarchical heavy neutrinos and
then an expression for the effective leptogenesis phase,
determining the conditions for maximal phase and placing a lower
bound on the phase suppression for generic models. Reconsidering
the lower bounds on the lightest right-handed (RH) neutrino mass
$M_1$ and on the reheating temperature $T_{\rm reh}$, we find that
models where one of the two heavier neutrino masses is dominated by
the lightest right-handed (RH) neutrinos, typically
arising from connections with quark masses, undergo both phase
suppression and strong wash-out such that $M_1 (T_{\rm reh})\gtrsim
10^{11}\,(10^{10})\,{\rm GeV}$. The window
$10^9\,{\rm GeV}\lesssim M_1,T_{\rm reh}\lesssim 10^{10}\,{\rm
GeV}$ is accessible only for a class of models where $m_1$ is
dominated by the lightest RH neutrino, with no straightforward
connections with
quark masses. Within this class we describe a new scenario of
thermal leptogenesis where the baryon asymmetry of the Universe is
generated by the decays of the second lightest RH neutrino, such
that the lower bound on $M_1$ disappears and is replaced by a
lower bound on $M_2$. Interestingly, the final asymmetry is
independent on the initial conditions. We also discuss the
validity of the approximation of hierarchical heavy neutrinos in a
simple analytical way.

\end{abstract}

\newpage
\section{Introduction}

The seesaw mechanism \cite{seesaw} is an elegant theoretical explanation for the lightness of
neutrinos compared to the other Standard Model particles.
Neutrino masses are predicted to be small but non-vanishing
and this has been successfully confirmed in neutrino mixing experiments \cite{SK}.
Moreover, within the seesaw, neutrino mixing data point to a new scale
$\sim M^2_{EW}/(\sqrt{\D m^2_{\rm sol}}-\sqrt{\D m^2_{\rm atm}})\sim 10^{14-15}\,{\rm GeV}$,
compatible with GUT's expectations \cite{GUTs}.

Beyond neutrino mixing and masses, the seesaw mechanism has
different phenomenological implications with interesting
testable predictions. An intriguing cosmological consequence of the seesaw mechanism
is  leptogenesis \cite{fy}. This provides an attractive explanation
for the observed baryon asymmetry of the Universe.
Analogously to the case of GUT baryogenesis models \cite{kt},
the asymmetry is generated from heavy particle decays. However,
in the case of leptogenesis, a B-L asymmetry is generated in the form of
lepton number that is
partly converted into a baryon asymmetry by (B-L conserving) sphaleron processes
\cite{sphaleron}. In a {\em thermal scenario} the initial temperature
of the Universe has to be high enough
to assure a sufficient heavy particle production and their kinematic equilibrium.
It is then quite remarkable that in this case neutrino mixing data favour
a simple picture of leptogenesis where just decays and inverse decays
provide a very good approximation \cite{bdp4} and the light neutrino
spectrum is hierarchical, with a stringent upper bound on the
absolute neutrino mass scale $m_i < 0.1\,{\rm eV}$
\cite{bdp1,bdp2,bdp3}.
 The role of the heavy decaying particles is played by three
new heavy RH neutrinos whose existence is predicted by the seesaw mechanism
and whose mass matrix is related to the light neutrino mass matrix $m_{\nu}$
by the well known seesaw formula,
\begin{equation}\label{seesaw}
m_\n = - m_D {1\over M} m_D^T\; \, ,
\end{equation}
where $m_D=v\,h$ is the Dirac neutrino mass matrix generated by the
Yukawa coupling matrix $h$ and $v$ is the Higgs vacuum expectation value.


It is always possible to work in a basis where the heavy neutrino
mass matrix is diagonal, $M={\rm diag}(M_1,M_2,M_3)\equiv D_M$,
with $M_1\leq M_2 \leq M_3$.
Moreover
one can also simultaneously diagonalize the light neutrino
mass matrix $m_\n$ by mean of the unitary matrix $U$, such that
\be
U^{\dagger}\,m_\n\,U^{\star}=-D_m \, ,
\ee
where $D_m \equiv {\rm diag}(m_1,m_2,m_3)$ and $m_1\leq m_2\leq m_3$.
The unitary matrix $U$ can be identified
with the PMNS mixing matrix in the
basis where the charged lepton mass matrix is diagonal
\footnote{This is rigorously valid in the limit
where the heavy neutrinos are infinitely heavier than the
Dirac mass matrix eigenvalues and the seesaw formula is exact.}.
In this way the seesaw formula (\ref{seesaw}) gets specialized in the following way:
\be
D_m=U^{\dagger}\,m_D\,D_M^{-1}\,m_D^{T}\,U^{\star} \, .
\ee
This expression can be also re-casted as an orthogonality condition,
\be
\Omega\,\Omega^T=\Omega^T\,\Omega=I \, ,
\ee
for the $\Omega$ matrix defined as \cite{casas}
\be\label{Omega}
\Omega=D_m^{-1/2}\,U^{\dagger}\,m_D\,D_M^{-1/2}
\ee
and whose matrix elements are then simply given by
\be\label{Omegah}
\Omega_{ij}={\widetilde{m}_{Dij}\over \sqrt{m_i\,M_j}}=
{v\,\widetilde{h}_{ij}\over \sqrt{m_i\,M_j}}\, ,
\ee
where $\widetilde{m}_D=U^{\dagger}\,m_D$ and $\widetilde{h}=U^{\dagger}\,h$.
The orthogonal  matrix $\Omega$ is fully determined by three complex parameters.
Inverting the  relation (\ref{Omega}) one obtains
\be\label{mD}
m_D=U\,D_m^{1/2}\,\Omega\,D_M^{1/2} \, .
\ee
 Together with the 6 (real) parameters necessary to specify the
$\Omega$ matrix, the seesaw parameters are given by the three light neutrino
masses, the 3 heavy neutrino masses and the 6 parameters (3 mixing angles and
3 phases) that fix the mixing matrix $U$ \cite{LisboaSaclay}.
The expression (\ref{mD}) is particularly enlightening because
the `observable' quantities are grouped apart on the RH side, whereas, on the LH side,
$m_D$ can be regarded as the theoretical quantity to be specified
within some model or set of assumptions
\footnote{The 3 heavy neutrino masses are
fundamental quantities but, through leptogenesis, they can
be also regarded as direct `observable' quantities, in the general
meaning explained further on.}.
Among the 18 observables, we can actually measure directly only a few of
them.  A complementary useful information
can however be inferred when leptogenesis is assumed
to explain the matter-anti matter asymmetry of the Universe.

Neutrino mixing data provide two important piece of information
on the neutrino mass spectrum. In the case of {\em normal (inverted) hierarchy} one has
\bea
m^2_3-m^2_2 & = & \Delta m^2_{\rm atm}\,(\Delta m^2_{\rm sol}) \, , \\
m^2_{2}-m^2_1 & = & \Delta m^2_{\rm sol}\,(\Delta m^2_{\rm atm})\, .
\eea
The third undetermined degree of freedom, the {\em absolute neutrino mass scale},
can be conveniently expressed in terms of the lightest neutrino mass $m_1$.
 The two heavier neutrino masses are then given by
\begin{eqnarray}
\label{numanor1}
m_3^{\,2}
&=& m_1^2 + m_{\rm atm}^2\;, \\
\label{numanor2}
m_2^{\,2} &=& m_1^2 + m^2_{\rm sol}\,(\Delta m^2_{\rm atm}),
\end{eqnarray}
where we defined $m_{\rm atm}\equiv\sqrt{\Delta m^2_{\rm atm} + \Delta m^2_{\rm sol}}$
and $m_{\rm sol}\equiv\sqrt{\Delta m^2_{\rm sol}}$.
The latest measurements give, combining the results from accelerator
and atmospheric neutrino experiments \cite{atm},
\begin{equation}
\D m^2_{\rm atm}= (2.3\pm 0.4)\times 10^{-3}\,{\rm eV}^2\;,
\end{equation}
and from solar neutrinos experiments \cite{solar}
\begin{equation}
\D m^2_{\rm sol}\simeq  (8.2^{+0.6}_{-0.5} \times 10^{-5})\,{\rm eV^2}\; \, ,
\end{equation}
implying
\be
  m_{\rm atm}=(4.9\pm 0.4)\times 10^{-2}\,{\rm eV} \,
  \hspace{7mm} \mbox{and} \hspace{7mm}
  m_{\rm sol}=(9.0^{+0.3}_{-0.25})\times 10^{-3}\,{\rm eV} \, .
\ee
Neutrino mixing experiments also measure
two of the three mixing angles, $\theta_{12}$ and $\theta_{23}$,
and place an upper bound on $\theta_{13}$. In
future they should be able to place, or hopefully measure,
a stringent constraint on the Dirac phase. On the other hand
they are insensitive to the two Majorana phases but some information
can be extracted from neutrinoless double beta decay experiments.


It could seem that there is no hope to get information on the
other `high energy' parameters, the three heavy neutrino masses and the 6 parameters
necessary to describe the orthogonal $\O$ matrix. However, within leptogenesis,
the baryon asymmetry is explained as the relic trace of a high energy scale
dynamical process, whose final value depends also on those seesaw parameters that
escape conventional experimental investigation.

Cosmic Microwave Background observations measure the baryon
asymmetry of the Universe with a few per cent precision that
is further improved when the information from large
scale structure is included.
A recent measurement, expressed in terms of the baryon to photon number ratio
at the recombination time, is given by the  WMAP
\cite{WMAP} plus SLOAN \cite{SLOAN} experiments that find
\be\label{etaBCMB}
\eta_{B}^{\rm CMB}=(6.3\pm 0.3)\times 10^{-10} \, .
\ee
There are different ingredients that enter the calculation of the
final baryon asymmetry in leptogenesis. In principle the predicted
final asymmetry depends on all seesaw parameters, 
however the dependence
on many of them is marginal and in this way the information from leptogenesis,
that in general would spread out through all parameters, focuses on a restricted
subset on which useful constraints can be derived.
This marginal dependence on many of the parameters, as we will see
those associated to the asymmetry generation and wash-out from the
two heavier RH neutrinos,  follows not only from
intrinsic features of leptogenesis but also because neutrino mixing data,
the large mixing angles and the values of $m_{\rm atm}$ and $m_{\rm sol}$,
select regions in the space of parameters where
the final asymmetry depends on a limited number of seesaw parameters.
Moreover it is remarkable that, within leptogenesis, there is a nice
conspiracy between the observed baryon asymmetry and neutrino mixing data,
such that their compatibility is realized for a wide variety of models of
neutrino masses and mixing, without particular tuning. There is however
an important limitation represented by a lower limit on the lightest
heavy neutrino mass $M_1$ \cite{di02,bdp1} and a related one on the
initial temperature $T_{\rm in}$ of the radiation dominated Universe expansion
\cite{bdp4} that, within an inflationary picture, can be identified
with the reheating temperature $T_{\rm reh}$ \cite{gkr}.

In Section 2 we  outline the different steps and approximations
that lead to predictions of the baryon asymmetry depending only on a restricted number
of seesaw parameters. This will enlighten the importance of the
maximum $C\!P$ asymmetry and of the effective leptogenesis phase to describe such predictions.
We will concentrate our attention on these two quantities in the
subsequent analysis. A useful tool is provided by the `seesaw geometry'
that we introduce in Section 3 and consists in representing the
orthogonal $\Omega$ matrix in the complex plane. In this way, in Section 4,
we are able to determine the leptogenesis $C\!P$ bound,
generalizing and clarifying existing results \cite{di02,bdp3,hambye} and
specifying the conditions on the $\O$ matrix for the effective leptogenesis
phase to be maximal. In Section 5 we find an expression for the effective leptogenesis phase
placing an interesting lower bound
and studying the conditions on the $\O$ matrix parameters for
successful leptogenesis.
Our procedure is phenomenological,
without any assumption on the matrix of the Yukawa couplings, while
the compatibility between different theoretical predictions
on $m_D$ with the observed baryon asymmetry and neutrino mixing data
has been investigated  in many works \cite{many}.
On the other hand, we will often point out different interesting connections
between our results and neutrino models, in particular in connection
with the lower bounds on $M_1$ and on $T_{\rm reh}$.
In Section 6 we discuss a new scenario of thermal
leptogenesis where the final asymmetry is generated by the decays of
the second lightest RH neutrino. This is possible when the wash-out
from the lightest RH neutrinos is weak and we will see how this
new possibility solves the well known problems of this regime,
namely the dependence on the initial
conditions and the fine-tuning of the effective neutrino mass $\mt$.
In Section 7 we examine the conditions of validity of the assumption
of a hierarchical heavy neutrino spectrum.
In Section 8 we summarize and draw some final conclusions.

\section{Seesaw parameters and leptogenesis}

The final baryon asymmetry from leptogenesis is, in general, a function of all
seesaw parameters. More explicitly one has
\be
\eta_B=\eta_B(m_1,M_i,\theta_{13},\delta,\varphi_1,\varphi_2,\O;m_{\rm atm},
m_{\rm sol},\theta_{12},\theta_{23}) \, ,
\ee
where the parameters measured in neutrino mixing experiments
are indicated after  the semi-colon. Their values justify
some approximations and simplifications, such that
the final asymmetry depends on a restricted number of parameters \cite{bdp1}.

An important approximation is to neglect the
asymmetry generated by the decays of the two heavier RH neutrinos
calculating the final baryon asymmetry only from the decays
of the lightest ones. We will consider in Section 6 a special scenario
where this approximation does not hold
while in Section 7 we will specify the conditions for its validity.
On general grounds this approximation is well justified because neutrino
mixing data favor a picture where, even for a mild heavy neutrino mass hierarchy,
the asymmetry generated by the two heavier neutrinos
is efficiently washed-out and can be safely neglected.
This is true for two reasons.
The first is that large mixing angles produce a flavor misalignment such that
the asymmetry produced from the decays of the two heavier RH neutrinos can be
washed-out mainly by the inverse decays of lightest RH neutrinos and by other
secondary processes. The second is that, the experimental fact that
$m_{\rm atm},m_{\rm sol}\gg 10^{-3}\,{\rm eV}$, makes typically such a
wash-out highly efficient and only what is produced by the
lightest RH neutrino decays around the {\em baryogenesis temperature}
$T_B\ll M_1$ survives until the present. This also implies that one has to assume the
initial temperature to be slightly higher than $T_B$ \cite{bdp1,bdp4}.
With this assumption, one can more generally neglect any initial asymmetry
and assume an initial thermal abundance of the lightest RH neutrinos
without worrying about the exact initial conditions
and of a detailed description of the thermalization process \cite{bdp3,bdp4}.
In this case $\eta_B$ from leptogenesis can be calculated as
\be\label{etaB}
\eta_{B}={a_{\rm sph}\over N_{\gamma}^{\rm rec}}\;\ve_1\;\k_{\rm f}  \, .
\ee
The sphaleron converting coefficient $a_{\rm sph}\sim 1/3$ is the
fraction of the $B-L$ asymmetry that ends up as a baryon asymmetry because
of sphaleron conversion \cite{ksht}. The number of photons at recombination
(in the portion of comoving volume containing, on average,
one $N_1$ in ultra-relativistic equilibrium) can be calculated
assuming a standard thermal history and in this case
$N_{\gamma}^{\rm rec}=4\,(g_{SM}+7/4)/(3\,g_{\rm rec})\simeq 37$.

The {\em $C\!P$ asymmetry parameter} is defined
\footnote{To have it positive, we have introduced
a minus sign compared to the usual definition.}
as
\be
\varepsilon_1=-{\Gamma-\bar{\Gamma}\over \Gamma+\bar{\G}} \, ,
\ee
where $\Gamma$ ($\bar{\Gamma}$) is the decay rate into leptons
(anti-leptons) and it gives the $(B-L)$ asymmetry generated per single decay
on average.

The {\em final efficiency factor} $\k_{\rm f}$ takes into account both
the heavy neutrino production and the wash-out due to
different processes, in particular inverse decays. It is
defined in a way to be one in the asymptotic limit of vanishing decaying
rate, initial ultra-relativistic thermal neutrino abundance.
If the condition $M_1\,\sum_i\,m^2_i \ll 10^{14}\,{\rm GeV}\,m^2_{\rm atm}$
holds, then the efficiency factor depends only
on the {\em effective neutrino mass} $\widetilde{m}_1$, defined as \cite{plum}
\be
\mt\equiv {(m_D^{\dagger}\,m_D)_{11}\over  M_1} \, .
\ee
Plugging the relation (\ref{mD}) into
the effective neutrino mass definition, one easily gets \cite{hama}
\be\label{mt1}
\mt=m_1\,\rho_{11}+m_2\,\rho_{21}+m_3\,\rho_{31}  \, ,
\ee
where we have introduced the polar representation
\be\label{polar}
\O_{ij}^2 = \rho_{ij}\,e^{i\,\varphi_{ij}} \, ,
\ee
with $\rho_{ij}\equiv |\O_{ij}^2|\geq 0$.
Since the first row of the $\O$ matrix will
play a special role in our discussion, it is useful to simplify the notation
such that $\rho_{j1}\rightarrow \rho_j$ and
$\varphi_{j1}\rightarrow\varphi_j$.
The orthogonality of the $\Omega$ matrix implies that
\be\label{orth}
\O_{11}^{\,2}+\O_{21}^{\,2}+\O_{31}^{\,2}=1 \, ,
\ee
from which it follows that $\mt\geq m_1$ \cite{hama}. This
is the only fully model independent restriction on $\mt$.
For configurations such that
\be
\sum_j\,\rho_j \sim \sum_j\,\Omega_{j1}^2=1
\ee
one has $\mt\lesssim m_3$. Models with $\mt\gg m_3$ rely on
the possibility of fine tuned phase cancellations, since only in this
way one can have $\rho_{3}\gg 1$, with
$\rho_{2}$ or $\rho_{1}$, or both, comparable to $\rho_{3}$.
Independently on whether  these models can provide or not a satisfactory solution
to describe neutrino masses and mixing and whether they can or cannot be justified
within some general theory, one has also to consider the
leptogenesis upper bound $\mt\lesssim (0.009\,{\rm eV})\,(M_1/10^{10}\,{\rm GeV})^{0.8}$,
valid  for $M_1\ll 10^{14}\,{\rm GeV}$
\footnote{This bound can be simply obtained inverting the  lower bound
on $M_1$ given in Section 5 (cf. (\ref{lbM1}) and (\ref{kfsw})).
For $M_1\gtrsim 10^{14}\,{\rm GeV}$ the bound is even more restrictive.
The maximum possible value for $\mt$, independently on the $M_1$ value,
is obtained at $M_1\simeq 5\times 10^{13}\,{\rm GeV}$ and is about
$\mt\sim 10\,{\rm eV}$, anyway much higher than the naive
expectation, often quoted in the literature,
$\mt\lesssim m_{\star}\simeq 10^{-3}\,{\rm eV}$,
from the full out-of-equilibrium condition. This
bound holds under the conditions on $\ve_1$ studied in
Section 7.}.

Conversely, there are models
with $\rho_{i}\simeq 1 \gg \rho_{j\neq i}$. Since the values of
the ${\rm Re}(\O^2_{ij})$'s determine the contribution of the heavy neutrino $N_j$
to the determination of the light neutrino mass $m_i$ \cite{LisboaSaclay,masina},
these kind of models are characterized
by the dominance of the lightest RH neutrino
in the determination of the lightest neutrino mass $m_i$.
In the strict limit $\rho_{i}=1$ and $\rho_{j\neq i}=0$, one has
necessarily $\mt=m_i\leq m_3$. If the other two light neutrino
masses, $m_{j\neq i}$, are also dominantly determined, by one
of the two heavier RH neutrinos, and if ${\rm Im}(\O^2_{ij})$ vanish,
then one has a particularly simple sub-set of the class models
with {\em sequential RH neutrino dominance} \cite{kingrep}.
In the limit case of exact dominance, there are 6
possible choices for the $\O$ matrix given by,

\be\label{SD}
\left(
\begin{array}{ccc}
       1 & 0 & 0 \\
       0 & 1 & 0 \\
       0 & 0 & 1
\end{array}\right) \, ,
\left(
\begin{array}{ccc}
       0 & 1 & 0 \\
       1 & 0 & 0 \\
       0 & 0 & 1
\end{array}\right) \, ,
\left(
\begin{array}{ccc}
       0 & 0 & 1 \\
       0 & 1 & 0 \\
       1 & 0 & 0
\end{array}\right) \, ,
\ee
plus the three cases obtained by exchanging the second and third row.
Obviously for these exact limits, the $C\!P$ asymmetry
and the effective leptogenesis phase vanish. However, it is enough to add
small complex perturbations to have non vanishing $C\!P$ asymmetry and
we will see that this can be even maximal if one starts from the first
matrix in Eq. (\ref{SD}).
An important issue for leptogenesis is whether $\mt$ is larger or smaller than the
{\em equilibrium neutrino mass} $m_{\star}\simeq 10^{-3}\,{\rm eV}$, that sets the transition
between the {\em weak} (for $\mt\lesssim m_{\star}$) and the
{\em strong wash-out regime} (for $\mt\gg m_{\star}$). In the latter case
thermal leptogenesis can work at its best with no dependence on the initial
conditions and minimal theoretical uncertainties \cite{bdp4}.
The possibility to have $\mt\lesssim m_{\star}$ relies on the fulfilment
of a few conditions.  First of all, since $\mt\geq m_1$, one must
have $m_1\lesssim 10^{-3}\,{\rm eV}\ll m_{\rm sol} \ll m_{\rm atm}$
and thus it requires fully hierarchical neutrinos.
From Eq. (\ref{mt1}) it follows that
$\rho_2,\rho_3\lesssim m_{\rm \star}/m_{\rm sol},m_{\rm \star}/m_{\rm atm}\ll 1$
and then, because of the orthogonality of $\O$, that $\r_1\simeq 1$.
This selects models with $\O$ matrices that are
perturbations of the complex 23-rotation
\be\label{O23}
\O=
\left(
\begin{array}{ccc}
  1  &  0   & 0   \\
  0  & \O_{22}             & \sqrt{1-\O^2_{22}} \\
  0 &  -\sqrt{1-\O^2_{22}} & \O_{22}
\end{array}
\right) \,
\ee
and that can be obtained performing additional `small' rotations in the
planes $13$ and $12$. A specific realization  has been shown in \cite{bdp3}.
For these models one has
$m_1\simeq v^2\,|\widetilde{h}^2_{11}|/M_1$ and thus the requirement
$m_1\leq\mt \lesssim m_{\star}$ implies  $|\widetilde{h}_{11}|\lesssim \alpha \equiv
M_1\,m_{\star}/v^2 \simeq 10^{-3}\sqrt{M_1/10^{10}\,{\rm GeV}}$.
At the same time one has to require
$\rho_{2(3)}\ll m_{\star}/m_{\rm sol(atm)}\simeq 0.125 (0.02) $, for
normal hierarchy, and this is equivalent to have
$|\tilde{h}_{i1}|\ll \alpha $. On the other hand in the limit $\mt\rightarrow m_1$,
as we will see, $\ve_1$ vanishes like $1-m_1/\mt$ and thus one has to impose
also $\mt\gg m_1$ in order to maximize the asymmetry.
In this way one arrives at the
conditions $|\tilde{h}_{11}|\ll |\tilde{h}_{21}|+|\tilde{h}_{31}|\lesssim \alpha$.
There are no compelling reasons to exclude these kinds of models,
but it is clear that they require a good amount of fine tuning to be realized.
Note also that the request that the lightest neutrino mass is
dominantly determined by the lightest RH neutrino mass
is such that these models are excluded when one imposes the experimental requirement
of  large mixing angles  and the theoretical requirement that the neutrino
Yukawa matrix resembles hierarchical quark matrices \cite{kingrep}.
These difficulties are in addition to the
problem, within leptogenesis, of the dependence of the final asymmetry
on the initial conditions. However, in Section 6 we will discuss
a new scenario  where $\mt\lesssim m_{\star}$ but the problems of the
fine-tuning of $\mt$ and of the dependence on the initial conditions are solved.
{\em For the time being we will refer to the more general
case of strong wash-out regime for $\mt\gg m_{\star}$.}

Let us now consider the $C\!P$ asymmetry.
 A perturbative calculation from the interference between tree level
and vertex plus self energy one-loop diagrams yields \cite{CPas}
\be\label{eps1g}
\varepsilon_1 =  -{1\over 8\pi}
\sum_{i=2,3}\,{{\rm Im}\,\left[(h^{\dagger}\,h)^2_{1i}\right]\over
(h^{\dagger}\,h)_{11}} \,\times\,
\left[f_V\left({M^2_i\over M^2_1}\right)+f_S\left({M^2_i\over M^2_1}\right)\right] \, .
\ee
The function $f_V$, describing the vertex contribution, is given by
\be
f_V(x)=\sqrt{x}\,\left[1-(1+x)\,\ln\left({1+x\over x}\right)\right] \, ,
\ee
while the function $f_S$, describing the self-energy contribution, is given by
\be
f_S(x)={\sqrt{x}\over 1-x}  \, .
\ee
Notice that this last expression is valid only when the two
mass differences $M_i-M_1$ are larger than the difference of
the respective two decay rates.
In the limit $x\gg 1$, corresponding to have a RH neutrinos mass hierarchy
with $M^2_i \gg M^2_1$, one has
\be\label{flimit}
f_V(x)+f_S(x)\simeq -{3\over 2\,\sqrt{x}} \, .
\ee
In this limit the
expression (\ref{eps1g}) simplifies into \cite{fred}
\be\label{eps1h}
\varepsilon_1 \simeq  {3\,M_1\over 16\pi}
\,{{\rm Im}\,\left[(h^{\dagger}\,h\,M^{-1}\,h^T\,h^{\star})_{11}\right]
\over (h\,h^{\dagger})_{11}} \,
\ee
and in Section 7 we will study the conditions of its validity.
Replacing $h$ with $\Omega$ (cf. (\ref{Omegah})), one then obtains \cite{di02}
\be\label{eps1Om}
\varepsilon_1 \simeq \ve_1(M_1,m_1,\mt,\Omega^2_{j1}) \equiv
{3\over 16\pi}\,{M_1\,m_{\rm atm}\over v^2}\,
\beta(m_1,\mt,\Omega^2_{j1}) \, ,
\ee
where we have introduced the convenient dimensionless quantity
\be\label{beta}
\beta(m_1,\mt,\Omega^2_{j1})=
{\sum_j\,m^2_j\,{\rm Im}(\Omega^2_{j1})\over m_{\rm atm}\,\sum_j\,m_j\,|\Omega_{j1}^2|}\, .
\ee
We have now considered all the different ingredients that enter
the prediction of the final baryon asymmetry within the approximation
of a hierarchical heavy neutrino spectrum and from Eq. (\ref{etaB})
we can see that $\eta_B=\eta_B(M_1,m_1,\mt,\O_{j1}^{\,2})$ .

Notice that
if the value of $\mt$ is fixed, then a generic configuration of
three $\O_{j1}^{\,2}$, because of the orthogonality condition (\ref{orth}),
depends only on 3 independent parameters.
Therefore, the final asymmetry depends just on 6 unknown seesaw parameters.
This is true within the validity of the approximation of a hierarchical
neutrino spectrum that made possible to cancel out a dependence on
$M_2$, $M_3$ and on two parameters of the $\O$ matrix.
Except for the absolute neutrino mass scale $m_1$,
the other 5 parameters cannot be measured in terrestrial experiments.
In Section 6 we will study a situation where,
even though the heavy neutrino spectrum is still assumed to be
hierarchical, the dependence of the final asymmetry on the
second lightest RH neutrino  cannot be neglected and
additional parameters enter the prediction of the final asymmetry.
Conversely, in Section 7, we will study the conditions for the
validity of the assumption of hierarchical heavy neutrino spectrum
and where this breaks down, such that, again, additional parameters
have to be taken into account for the determination of the final asymmetry.

The efficiency factor $\k_{\rm f}$ depends only on $\mt$, $m_1$ and $M_1$, implying
that the dependence of the final asymmetry on the $\O_{j1}$'s arises only from $\ve_1$.
It is then quite interesting to search for those particular configurations
of the $\Omega_{j1}^2$'s such that, for fixed values of $m_1$ and $\mt$,
the $C\!P$ asymmetry is maximum and, for this purpose, it is useful to
define an {\em effective leptogenesis phase} $\d_L$ \cite{asaka} such that
\be\label{lepphase}
\beta(m_1,\mt,\Omega^2_{j1})=\beta_{\rm max}(m_1,\mt)\,\sin\d_L(m_1,\mt,\Omega_{j1}^2) \, .
\ee
The function $\beta_{\rm max}(m_1,\mt)\geq 0$ is the maximum value of $\b$
corresponding, when plugged into Eq. (\ref{eps1Om}), to
a maximum  $C\!P$ asymmetry $\ve_1^{\rm max}(M_1,m_1,\mt)\geq 0$,
realized for those particular configurations with $\sin\d_L=1$.
Notice that a trivial bound on the $C\!P$ asymmetry is,
by definition, $\ve_1\leq 1$. It is however possible to find a more
restrictive non trivial bound within the validity of Eq. (\ref{eps1h})
\cite{asaka,di02,bdp3,hambye}.
The {\em maximum baryon asymmetry} is given by
\be\label{etaBmax}
\eta_B^{\rm max}(M_1,m_1,\mt)=d\,\ve_1^{\rm max}(M_1,m_1,\mt)\,\k_{\rm f}(M_1,m_1,\mt) \, ,
\ee
where $d\equiv a_{\rm sph}/N_{\gamma}^{\rm rec}$ and it depends only on the
three seesaw parameters $\mt$, $m_1$ and $M_1$. Successful leptogenesis implies
$\eta_B^{\rm max}(M_1,m_1,\mt) \geq \eta_B^{CMB}$ and this yields
interesting leptogenesis constraints on $\mt,m_1$ and $M_1$
\cite{di02,bdp1} and in particular an
upper bound on the absolute neutrino mass scale $m_1$ \cite{bdp1,bdp2,bdp3}
testable in cosmological, neutrinoless double beta decay and
Tritium beta decay experiments.

For an accurate evaluation  of these constraints, it is thus
necessary to determine the function $\beta_{\rm max}(m_1,\mt)$
and it is important to understand all properties
to be fulfilled in order to saturate the bound. This means
a determination of the class of configurations of $\Omega_{j1}^{\,2}$'s
such that  $\sin\d_L=1$. At the same time
we will be interested in determining the phase $\sin\d_L$
for an arbitrary configuration of $\Omega_{j1}$'s and its connections with
neutrino models. In this way one can answer
interesting questions on the {\em stability of the $C\!P$ bound}, that means
whether small variations of the seesaw parameters determine a small or a large
variation of the leptogenesis phase.
This will make possible to realize whether the condition of maximal phase
should be regarded as a reasonable or a fine tuned assumption. Moreover,
placing a lower bound on the same leptogenesis phase will provide
interesting information on the geometrical parameters.

\section{Seesaw geometry}

Let us represent the three $\Omega_{j1}^2$'s in the complex plane. The orthogonality
condition fixes the sum of the three to start from the origin and to
end up onto the real axis at the point
\be
{\rm Re}\, \sum_j\Omega^2_{j1}=1 \, ,
\ee
as shown in Fig. 1 for a generic configuration (dashed line arrows).
\begin{figure}[t]
\hspace{-5mm}
\centerline{\psfig{file=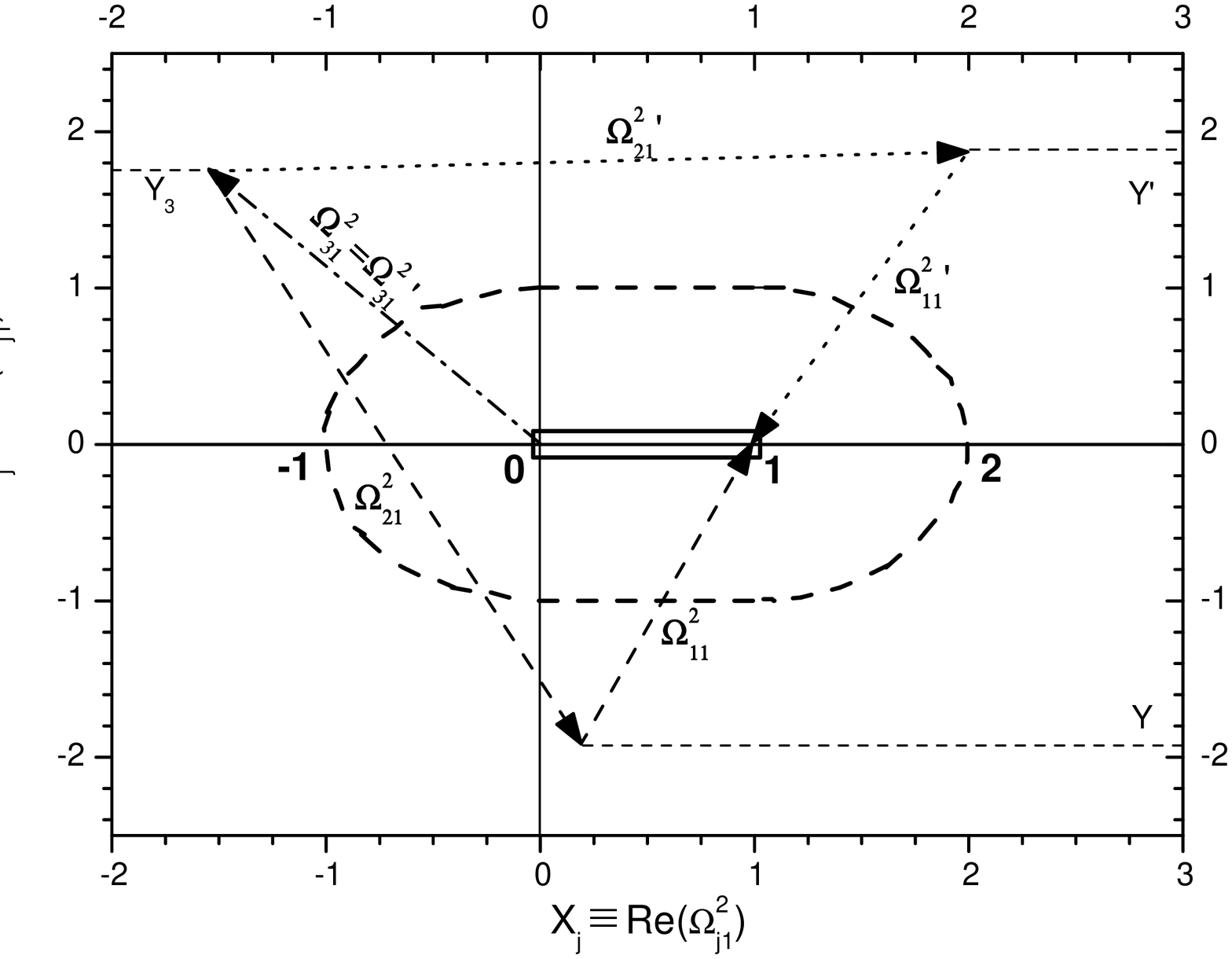,height=12cm,width=115mm}}
\vspace{-10mm} \caption{\small seesaw geometry. An arbitrary
$\O^2_{j1}$ configuration with ${\rm sign}(Y_3)=-{\rm sign}(Y)$
(dashed line arrows) and its  $\O^{2\,'}_{j1}$ `dual' one with
${\rm sign}(Y_3)={\rm sign}(Y)>0$ (dotted line arrows). The region
inside the short-dashed line is that one spanned by the areas
delimited by the configurations with $|\O_{j1}^{\,2}|\equiv\rho_j
\leq 1$ and the segment $[0,1]$. Configurations corresponding to
models with $\r_i\simeq 1 \gg \r_{j\neq i}$ lie within a region,
schematically indicated with the box, around the $[0,1]$ segment
on the real axis.}
\end{figure}
In addition to the polar representation (cf. (\ref{polar})), it is also
useful to indicate explicitly the real and the imaginary part
of the $\Omega_{j1}^2$'s, writing
\be
\Omega_{j1}^2\equiv X_j+i\,Y_j .
\ee
If one defines that class of models such that $\rho_{j}\leq 1$, then
the area delimited by the three $\O_{j1}^2$'s  and the segment $[0,1]$ is
confined to lie within a well defined region
in the complex plane inside the dashed line in Fig. 1.
For these models one has (cf. (\ref{mt1}))
\be\label{mt<}
\mt \leq m_1+m_2+m_3 \simeq
\left\{\begin{array}{ccc}
 m_{\rm atm}\;\; (\mbox{\rm fully hierarchical neutrinos}) \\
        3\,m_1 \;\;(\mbox{\rm quasi-degenerate neutrinos})
\end{array}\right. \, .
\ee
Models with $\rho_i\simeq 1 \gg \r_{j\neq i}$
correspond to all those geometrical configurations
with small deviations from the segment $[0,1]$ along the real
axis (see the box in Fig. 1). In particular, as we have already seen
in the previous section, only for models with
$\r_1\simeq 1\gg \r_2,\r_3$ (cf. (\ref{O23}))
one can have $\mt\lesssim m_{\star}$.

Conversely, the possibility of having $\mt\gg m_3$ relies on models with
strong phase cancellations with no dominance of just one $\rho_{j}$,
while at least two of them have to be comparable and much larger than one.
The two (dual) configurations depicted in Fig. 1, are an example of such a possibility.

An interesting aspect of this geometrical representation
is a comparison with the $C\!P$ violation in the weak interactions of quarks, described by the
CKM matrix, and in neutrino mixing, described by the PMNS matrix.
 In the quark sector a geometrical representation of $C\!P$ violation and of
the CKM matrix is provided by the unitarity triangles, whose area is given by
the Jarlskog parameter \cite{Robert}. If the triangles squeeze to a segment, then $C\!P$
violation vanishes. This is an easy way to understand why
a non vanishing $C\!P$ violation in the quark sector and in neutrino
mixing requires at least three generations. On the other hand in the case of the
$C\!P$ violation governed by the orthogonal $\O$ matrix
two generations are enough, since instead of a triangle one has,
in general, a quadrilateral where one side, the real segment $[0,1]$,
is present anyway. Thus, within the seesaw mechanism, one has
a kind of $C\!P$ violation that is different from the usual
one in the quark sector (or from the possible one in neutrino mixing),
since in the first case this is described by an orthogonal matrix,
while in the second case it is described by a unitary matrix.

\section{$C\!P$ asymmetry bound}

Let us now study the leptogenesis $C\!P$ asymmetry bound
taking advantage of the `seesaw geometry'.
Using the $\mt$ definition (cf. (\ref{mt1}))
 and the orthogonality condition (cf. (\ref{orth})),
 the function $\beta(m_1,\mt,\Omega^2_{j1})$,
defined by the Eq. (\ref{beta}), can be re-casted as
\be\label{bY}
\beta(m_1,\mt,\Omega^2_{j1})=
{\Delta m^2_{32}\,Y_3+\Delta m^2_{21}\,Y\over m_{\rm atm}\,\mt} \, ,
\ee
with $Y\equiv Y_2+Y_3$. Notice that from any configuration such that
${\rm sign}(Y)=-{\rm sign}(Y_3)$, one can always obtain two configurations with
equal signs, ${\rm sign}(Y)={\rm sign}(Y_3)=\pm 1$, same value
of $\mt$ and such that
$|\b(m_1,\mt,\Omega^2_{j1})|$ is higher. Since the observed baryon asymmetry
is conventionally defined to be positive, then the $C\!P$ asymmetry, the function $\beta$
and the effective leptogenesis phase $\delta_L$ have to be positive too
and thus we will always refer, in the following,
to configurations with positive $Y$ and $Y_3$ (an example is shown in Fig. 1).

It is easy to find the upper bound $\beta_{\rm max}(m_1,\mt)$ (cf. (\ref{lepphase}))
with $\mt$ free to change while $m_1$ is fixed. If we write $\mt$ as
\be\label{mt1XY}
\mt=m_1\,\sqrt{Y^2+(1-X)^2}+m_2\,\sqrt{X_2^2+Y_2^2}+m_3\,\sqrt{X_3^2+Y_3^2} \, ,
\ee
then $\b(m_1,\mt,\Omega^2_{j1})$ is maximized in the limits $Y_3\gg X_3$,
$Y_2\gg X_2$ and $Y\gg 1$, implying $\mt\gg m_1$, and
it gets further maximized for $Y_2=0$.
In this way one obtains the upper bound \cite{di02}
$\beta_{\rm max}(m_1,\mt)=
{(m_3-m_1)/ m_{\rm atm}}$,
corresponding to (cf. (\ref{eps1Om}))
\be\label{di2}
\ve_1(M_1,m_1,\mt,\O_{j1}^{\,2}) \leq \ve_{\rm max}(M_1)\,{m_3-m_1\over m_{\rm atm}} \, ,
\ee
where we defined
\be\label{e1maxM1}
\ve_{\rm max}(M_1)\equiv {3\over 16\pi}\,{M_1\,m_{\rm atm}\over v^2}
\simeq 10^{-6}\,\left({M_1\over 10^{10}\,{\rm GeV}}\right)
\,\left({m_{\rm atm}\over 0.05\,{\rm eV}}\right) \, ,
\ee
giving the absolute maximum of the $C\!P$ asymmetry in the limit
of hierarchical neutrinos \cite{asaka,di02} for $m_1/m_{\rm atm}\rightarrow 0$.
For a given finite value of $\mt$ there is a more restrictive upper
bound that
in general, introducing a function $f(m_1,\mt)\leq 1$,
can be written as \cite{bdp3}
\be\label{f}
\beta_{\rm max}(m_1,\mt)={m_3-m_1\over m_{\rm atm}}\,\,f(m_1,\mt) \leq 1 \, ,
\ee
corresponding to
\be\label{e1max}
\ve_1(M_1,m_1,\mt,\O_{j1}^{\,2}) \leq \ve_{\rm max}(M_1)\,
{m_3-m_1\over m_{\rm atm}}\,\,f(m_1,\mt) \, .
\ee
For $\mt/m_1\rightarrow\infty$ one has
$f(m_1,\infty)=1$ and the limit (\ref{di2}) is recovered.

We have now to calculate the function $f(m_1,\mt)$ for
finite  values of $\mt$. Before dealing with the general case, for arbitrary $m_1$,
let us consider the two interesting asymptotic limits of {\em fully hierarchical neutrinos},
for $(m_1/m_{\rm sol})^2\rightarrow 0 $, and of {\em quasi-degenerate neutrinos},
for $(m_1/m_{\rm atm})^2\rightarrow\infty$.

\subsection{Fully hierarchical neutrinos}

In this case $\beta_{\rm max}=1$ and there is no global
suppression. Moreover one has
\be\label{mtfh}
\mt=m_2\,\rho_2+m_3\,\rho_3 \, ,
\ee
with $m_3=m_{\rm atm}$ and
$m_2=m_{\rm sol}\,\,(\sqrt{m_{\rm atm}^2-m_{\rm sol}^2})$
in the case of normal (inverted) hierarchy.
This implies that, for any change of configuration such that $\rho_2$
and $\rho_3$ remain constant, the quantity $\mt$ is constant too,
while  $\rho_1$ can be arbitrarily modified. Hence, for
a given value of $\mt$, one has that $\beta(0,\mt,\Omega_{j1}^2)$
is maximum for a configuration with $X_2=X_3=0$, such that
the Eq. (\ref{mtfh}) gets specialized into $\mt=m_2\,Y_2+m_3\,Y_3$\,.
Using this expression one can replace $Y_3$ into the Eq. (\ref{bY}), getting
\be
\b_{\rm max}(0,\mt)=f(0,\mt)={\rm max}\left[1-{m_2\over\mt}\,\left(1-{m_2\over m_3}\right)\,Y_2\right] \, ,
\ee
that is clearly maximized for $Y_2=0$,
corresponding to $Y_3=Y$ and $\mt=m_3\,Y_3$.
 Therefore, one has very simply $f(m_1=0,\mt)=1$ showing that
 the case $m_1=0$ yields, for a fixed $M_1$, an absolute
maximum of the $C\!P$ asymmetry given by $\ve_{\rm max}(M_1)$ (cf. (\ref{e1maxM1})),
as we have already found as a particular case of $\mt/m_1\gg 1$.
In Fig. 2 we show a generic configuration with a given $\mt$ (solid line arrows)
and the corresponding configuration $X_3=Y_3=Y_2=0$, with the same value of $\mt$,
for which the $C\!P$ asymmetry is maximum (dashed line arrows). It is worth
to notice that {\em this class of configurations correspond to the case when
the lightest neutrino mass is dominated by the lightest RH neutrino mass and
thus for any other kind of model there will be necessarily some phase suppression}
\footnote{In \cite{hk} it was found that, within single RH neutrino dominated models,
where the largest neutrino mass $m_3$ is dominated by one RH neutrino,
the  $C\!P$ asymmetry $\ve_1$ is larger when the largest neutrino mass $m_3$
is dominated by the heaviest RH neutrino compared to when is dominated by the lightest.
This can be regarded as a particular case of our more general result: if the lightest
RH neutrino dominates $m_3$ then it cannot dominate $m_1$, because of the
$\O$ orthogonality, and thus there must be a phase suppression. On the other hand
having $m_3$ dominated by the heaviest RH neutrino is compatible with
$m_1$ dominated by the lightest RH neutrino. The point is that
the RH neutrino dominance directly relevant for leptogenesis
is that of $m_1$ and not that of $m_3$.}.
\begin{figure}[t]
\centerline{\psfig{file=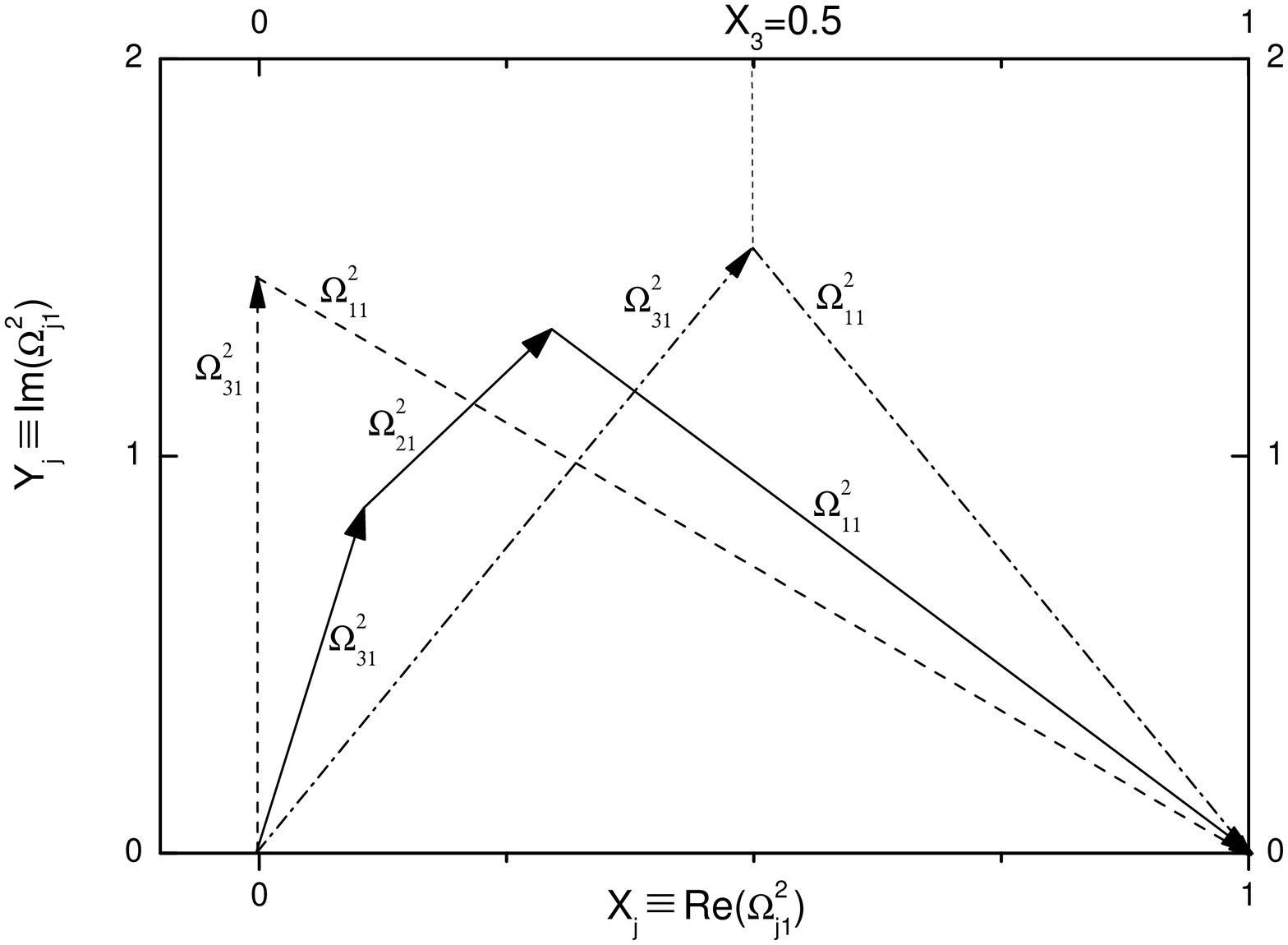,height=12cm,width=16cm}}
\vspace{-10mm}
\caption{\small An arbitrary configuration for a
given $\mt$ value (solid line arrows)
and the associated ones that, for the same $\mt$ value,
maximize the $C\!P$ asymmetry in the limits of
{\em fully hierarchical neutrinos} (dashed line arrows) and
{\em degenerate neutrinos} (dot-dashed line arrows).}
\end{figure}

\subsection{Quasi-degenerate neutrinos}

In the {\em quasi degenerate limit}, for $(m_1/m_{\rm atm})^2 \gg 1$,
one has $m_1 \simeq m_2 \simeq m_3$ and the expression for $\mt$ (cf. (\ref{mt1}))
becomes simply $\mt\simeq m_1\,{\ell}$,
with ${\ell}\equiv \sum_j\,\rho_j$. Therefore, the condition $\mt={\rm const}$
is equivalent to select all those configurations for which ${\ell}$ is constant.
 Then it simply turns out that
 $\beta(m_1,\mt,\Omega^2_{j1})$ is maximum for a configuration
such that $Y_2=X_2=0$ and $X_3=1/2$, as shown in Fig. 2 (dashed line arrows),
for the simple reason that only for this configuration both $Y_3$ and $Y$
are simultaneously equal to their maximum value. This is easy  to be understood
in a geometrical way: it corresponds to `stretch' the string with length
${\ell}$ as high as possible in order to maximize $Y$, while at the same time
the value of $Y_3$ is maximized by having $X_2=Y_2=0$.
 Since in the quasi-degenerate limit (neglecting terms ${\cal O}(m_{\rm atm}/m_1)^2$),
one has $m^2_{\rm atm}\simeq 2\,m_1\,(m_3-m_1)$, it is easy to obtain the
result (cf. (\ref{bY}) and (\ref{f}))
\be\label{qd}
f(m_1,\mt)=\sqrt{1-{m_1^2\over\mt^2}}  \,\,\, ,
\ee
already obtained in \cite{hambye} with a different procedure and using the approximation $m_{\rm sol}=0$.

\subsection{The general case}

Let us now find the configuration that maximizes
the $C\!P$ asymmetry and calculate $f(m_1,\mt)$
for arbitrary values of $m_1$. Notice that in both the two asymptotic
limits we have found $\O_{21}^{\,2}=0$ and so one can wonder
whether this condition holds in general for any value of $m_1$.
This can be proved showing that starting from a generic configuration with
$\rho_2\neq 0$ one can always find a configuration with the
same value of $\mt$, such that $\rho_2$ is lower and $\beta$
is higher. It is useful to split the demonstration in two steps.

The {\em first step} consists in showing that starting from a
generic configuration one can always find a configuration,
with higher $\beta$ and same value of $\mt$ and
such that $Y_2=0$ (i.e. $Y=Y_3$). The procedure to build such a
configuration is depicted in Fig. 3 and in Fig. 4, while we
address to the Appendix A for more details.
\begin{figure}[t]
\centerline{\psfig{file=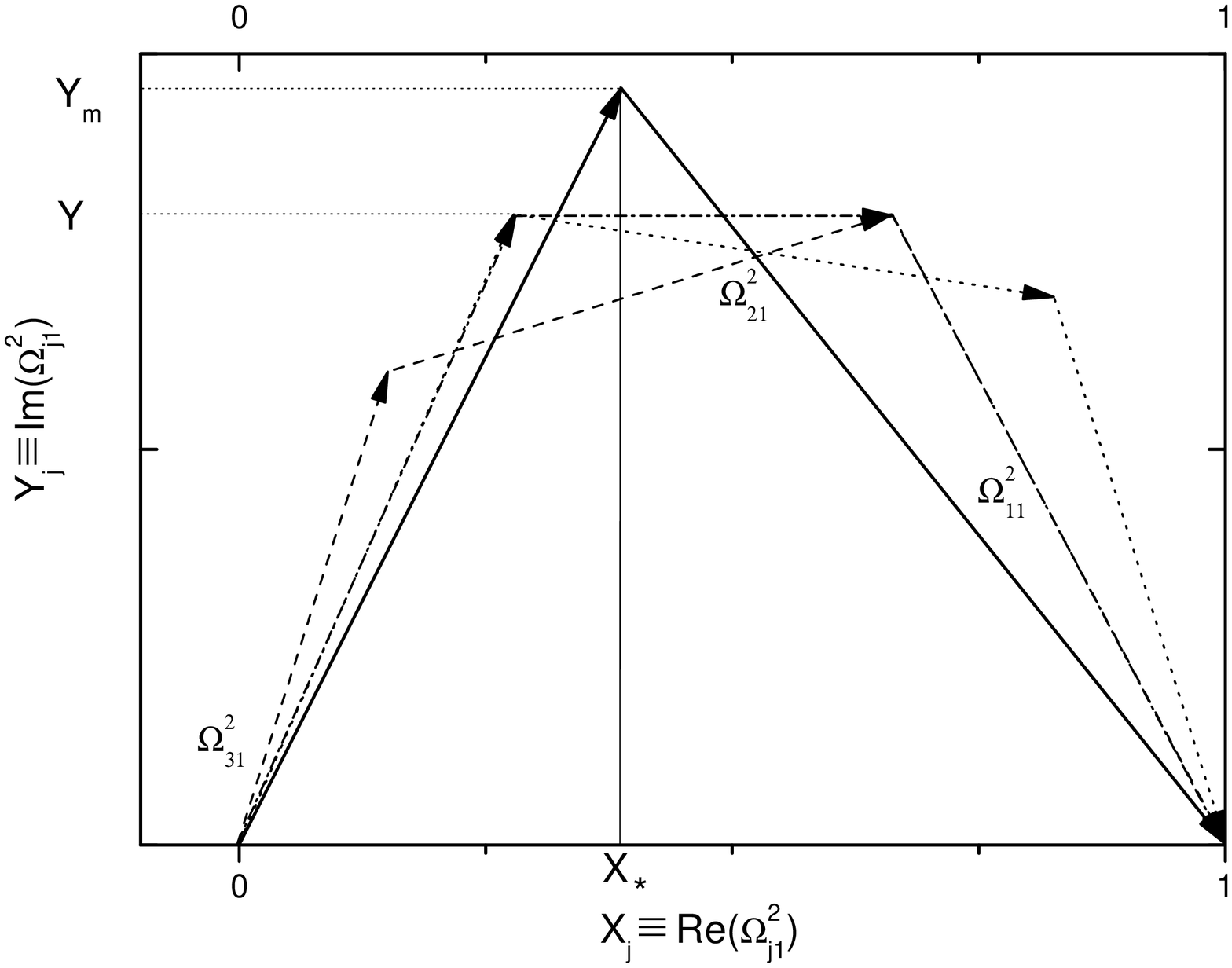,height=12cm,width=16cm}}
\vspace{-20mm}
\caption{\small Procedure that brings from a generic
configuration with $Y_2>0$ (dashed arrows) or $Y_2<0$
(dotted arrows) to one with $Y_2=0$ (dash-dotted arrows)
and from this to one with $\O^2_{21}=0$ (solid arrows) such that $\mt$ does not
change and the $C\!P$ asymmetry increases.}
\end{figure}
\begin{figure}[t]
\centerline{\psfig{file=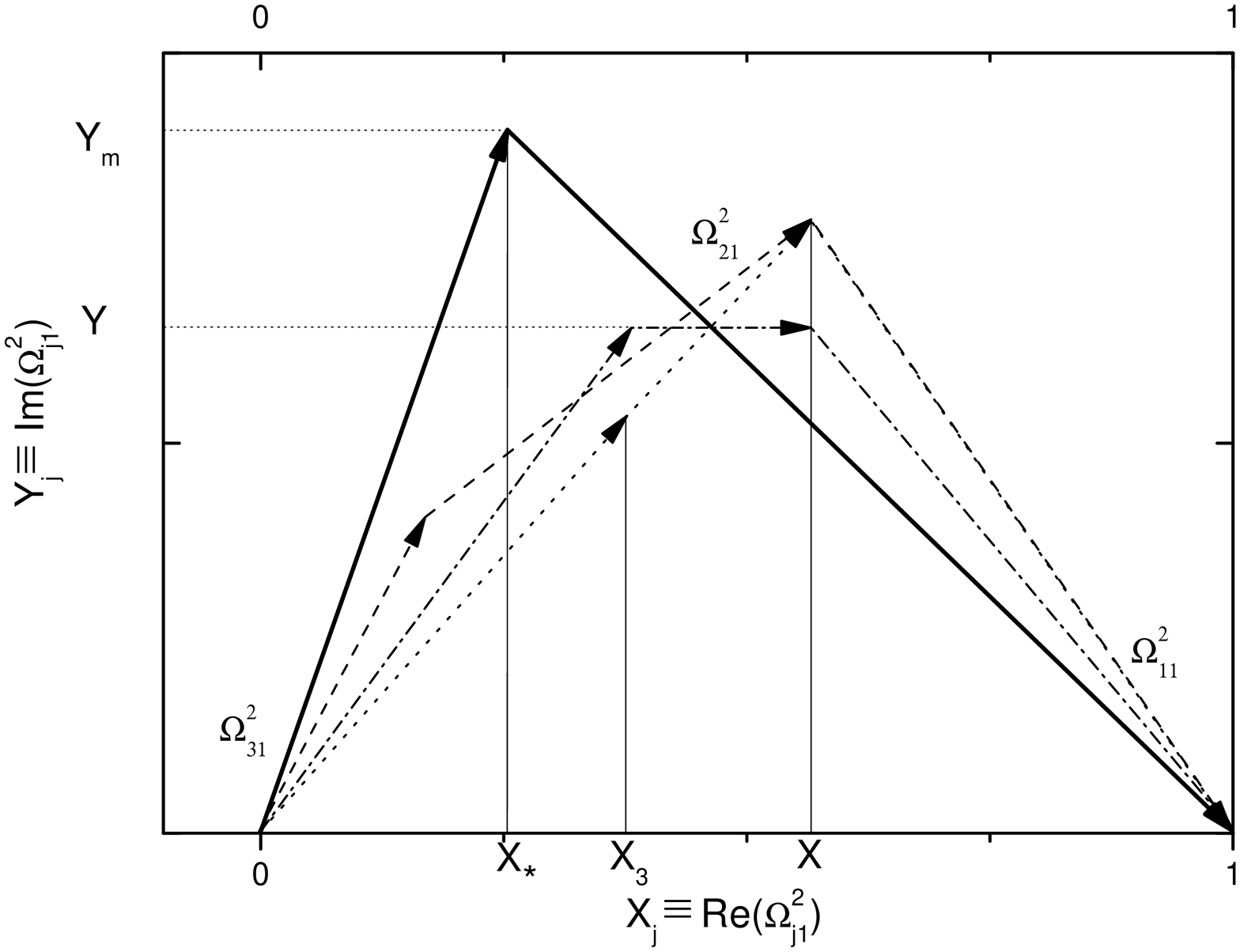,height=12cm,width=16cm}}
\vspace{-20mm}
\caption{\small An example of the procedure that
brings to a configuration with maximum asymmetry starting from
one with $Y_2>0$ (dashed arrows). In this example by decreasing $\rho_2$
the configuration gets first transformed into one with $\varphi_2=\varphi_3$
(dotted arrows). From this recursively, as explained in the Appendix,
one still ends up with a configuration with $Y_2=0$ (dot-dashed arrows)
and from this again to one with $\O_{21}^2=0$ (solid arrows).}
\end{figure}
The {\em second step} consists in showing that within the class of configurations
with a given value of $\mt$ and with $Y_2=0$, the value of
$\beta$ is further maximized when $X_2=0$ and thus, in conclusion,
when $\Omega^2_{21}=0$. Looking at the two configurations
in Fig. 3 and in Fig. 4 with $Y_2=0$ (dash-dotted line arrows),
it is quite intuitive geometrically that $Y$ increases when $X_2$
decreases and this is confirmed
by an analytic calculation. Indeed, the expression (\ref{bY}) simplifies into
\be\label{bYY20}
\beta(m_1,\mt,\Omega_{j1}^2)={m_{\rm atm}\over \mt}\,Y \, ,
\ee
where now $Y=Y(\mt,X_3,X_2)$ is a function of just three parameters.
Since $\mt$ is fixed, then $\beta$ is maximum when $Y$ is maximum.
In the {\em quasi-degenerate} case it is easy to understand that
$Y$ is maximum when $X_3=(1-X_2)/2$ and is further maximized when
$X_2=0$ such that $X_3=1/2$, confirming the result that we previously obtained
(cf. subsection 3.2) in another way. In the {\em hierarchical case}
($m_1=0$) it is simple to obtain from the Eq. (\ref{mtfh}) that
\be
Y^2={\mt\over m_3}-{m_2\over m_3}\,X_2^2-X_3^2,
\ee
that is maximum for $X_3=X_2=0$, confirming again what we have
previously found in subsection 3.1. In the  general case,
starting from the definition of $\mt$ (cf. (\ref{mt1XY})) with
$Y=Y_3$ and taking the derivative with respect to $\rho_2$,
one can easily show that $\partial Y/\partial \rho_2< 0$.
Thus the maximum value of $Y$ is obtained when $\O_{21}^2=0$.
Two examples of such configurations are shown in Fig. 3  and in Fig. 4
(solid arrows).
{\em We can thus conclude that for a given value of $m_1$ and $\mt$, the $C\!P$
asymmetry is maximized for configurations with $\O^2_{21}=0$.}


For a given value of $\mt$, these configurations are described just by one parameter,
and we can conveniently choose  $X=1-X_1=X_3$. Therefore,
the expression (\ref{mt1XY}) for $\mt$ can be re-casted as
\be\label{mtmax}
\mt=m_1\,\sqrt{Y^2+(1-X)^2}+m_3\,\sqrt{Y^2+X^2} \, .
\ee
The last step is thus to maximize the $C\!P$ asymmetry and this means to find
the maximum value of $Y$ entering the expression for $\b$ (cf. (\ref{bYY20})),
with respect to $X$. In this way one can
obtain the $C\!P$ asymmetry bound for given, arbitrary, values of
$\mt$ and $m_1$, not depending on the geometrical parameters $\O^2_{j1}$.
This can be found just
taking the derivative of the expression (\ref{mtmax}) with respect to $X$
and imposing $(dY/dX)_{X=X_{\star}}=0$. Doing this, one arrives
to the simple relation $m_3\cos\varphi_3=m_1\cos\varphi_1$, or explicitly
in terms of $X_{\star}$ and $Y_{\rm m}$
\be\label{m3m1}
{m_3\,X_{\star}\over
\sqrt{X_{\star}^2+Y_{\rm m}^2}}=
{m_1\,(1-X_{\star})\over \sqrt{(1-X_{\star})^2+Y_{\rm m}^2}} \, .
\ee
This expression, together with the Eq. (\ref{mtmax}) for $\mt$,
defines the functions $X_{\star}(m_1,\mt)$ and $Y_{\rm m}(m_1,\mt)$, such that
$\beta$ and thus the  $C\!P$ asymmetry are maximum.
Notice that
the function $f(m_1,\mt)$ (cf. Eq. (\ref{f}))
is related to $Y_{\rm m}(m_1,\mt)$ by
\be\label{fY}
f(m_1,\mt)=
{m_3+m_1\over \mt}\, Y_{\rm m}(m_1,\mt)
\, .
\ee
In Fig. 5 we have plotted $f$ and $X_{\star}$ vs.  $x\equiv m_1/\mt$
and for different values of $m_1/m_{\rm atm}$.
\begin{figure}[t]
\centerline{\psfig{file=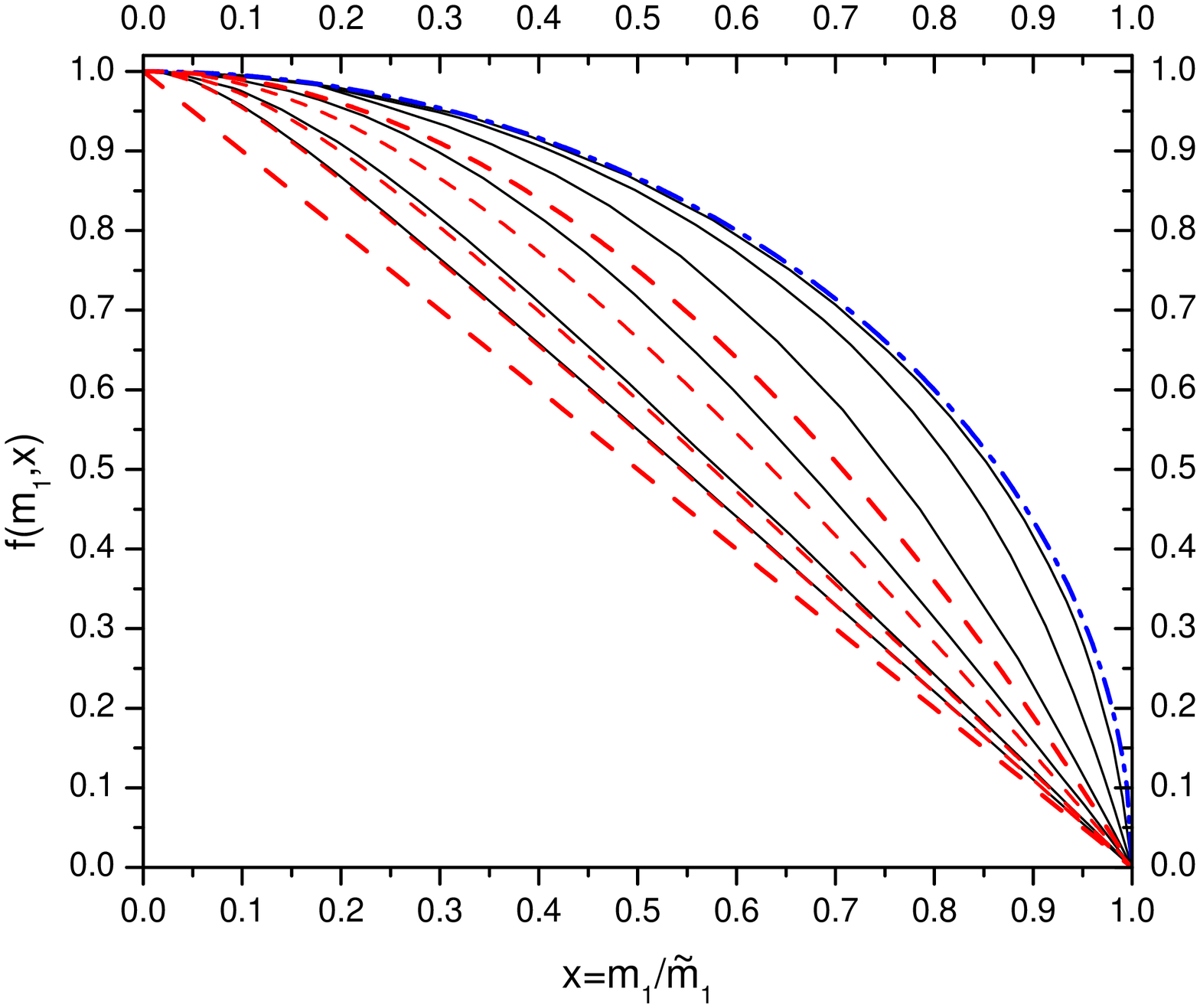,height=12cm,width=14cm}}
\vspace{-10mm}
\caption{\small function $f$ calculated from the Eq.'s
(\ref{mtmax}), (\ref{m3m1}) and (\ref{fY}) and
plotted vs. $x\equiv m_1/\mt$ for $m_1/m_{\rm atm}=5,2,1,0.5,0.2,0.1$
(from the top to the bottom solid line).
The hierarchical limit for $f$ (cf. (\ref{attempt})) is plotted
for $m_1/m_{\rm atm}=1,0.5,0.2,0.1,0.01$ (from the top to the bottom dashed line),
while the quasi-degenerate limit (cf. Eq. (\ref{qd})) corresponds to the
dot-dashed line.}
\end{figure}
%
%
\begin{figure}[t]
\centerline{\psfig{file=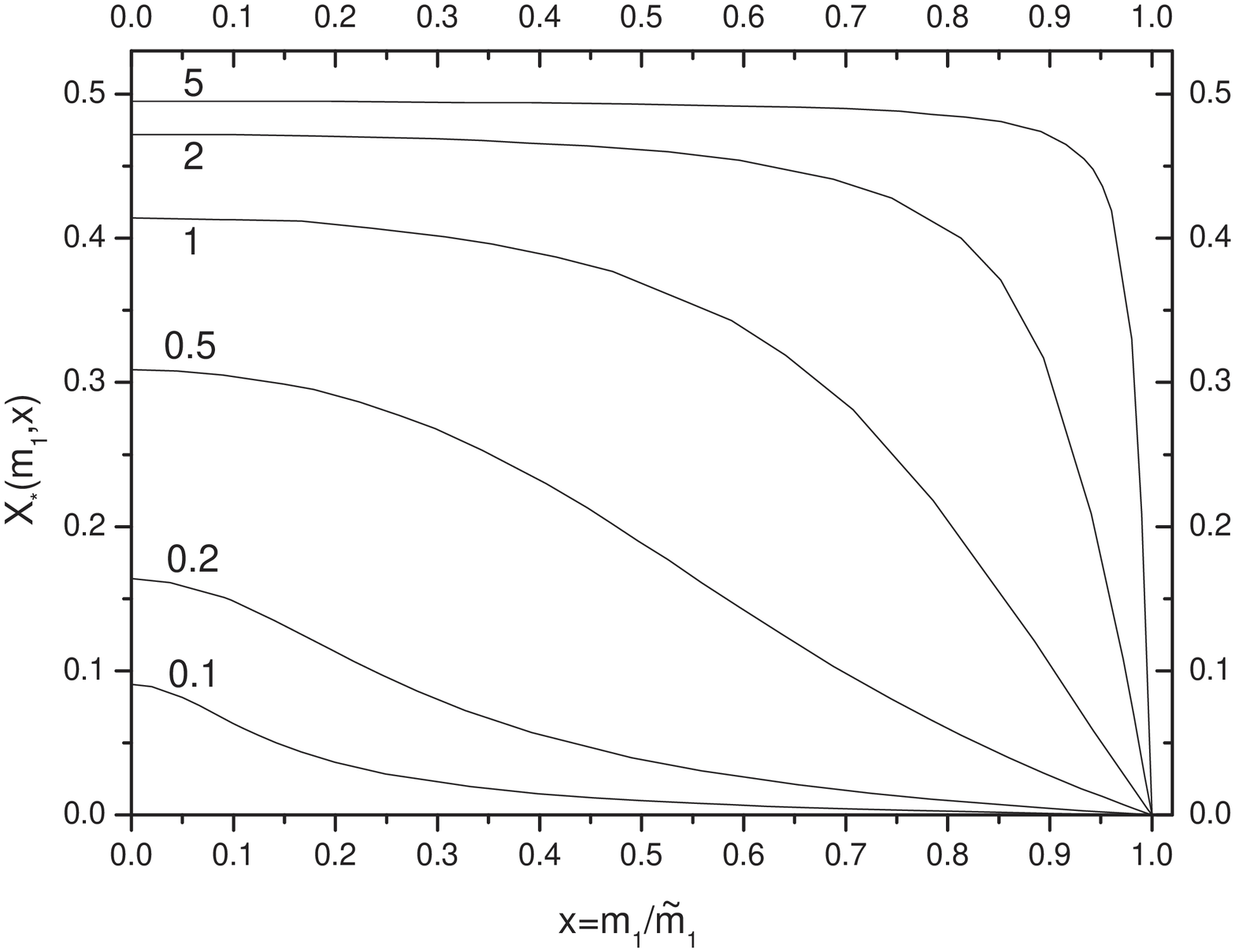,height=12cm,width=14cm}}
\vspace{-10mm}
\caption{\small $X_{\star}$ calculated from the Eq.'s (\ref{mtmax}) and (\ref{m3m1})
as a function of $x\equiv m_1/\mt$ and for the indicated values of $m_1/m_{\rm atm}$.}
\end{figure}
It is also possible to derive explicitly analytic expressions for the functions
$X_{\star}$ and $Y_{\rm m}$. However these are algebraically quite involved
and not useful to be written down, while the two equations
(\ref{mtmax}) and (\ref{m3m1})
provide a much more useful physical insight and at the same time
simple and useful explicit analytical expressions
can be easily derived in some interesting limits:
\begin{itemize}
\item in the {\em limit of fully hierarchical neutrinos} one obtains
$X_{\star}=0$ and $Y_{\rm m}=\mt/m_3$, implying the already well known
result $f(m_1,\mt)=1$;
\item in the {\em limit of quasi-degenerate
neutrinos}, for $m_1\simeq m_3$, one gets from the Eq. (\ref{m3m1})
$X_{\star}=1/2$, that inserted in the Eq. (\ref{mtmax}) gives
once more $f(m_1,\mt)=\sqrt{1-(m_1/\mt)^2}$;
\item in general, since $0\leq m_1/m_3 \leq 1$, one has $0\leq X_{\star}\leq 1/2$;
\item if $\mt=m_1$ 
then $X=Y=f=0$ and vice versa;
\item in the limit $\mt\gg m_1$ 
one gets easily
$Y_{\rm m}(m_1,\mt\gg m_1)=\mt/(m_1+m_3)$, 
corresponding to the already known result $f(m_1,\mt\gg m_1)=1$,
and also $X_{\star}(m_1,\mt\gg m_1)=m_1/(m_1+m_3)$. 
\end{itemize}
It is also easy to go beyond the limit of fully hierarchical neutrinos and find
an expression for the function $f(m_1,\mt)$ valid in the regime
$m_1/m_{\rm atm}\lesssim 1$, corresponding to $m_1\lesssim m_3/\sqrt{2}$.
The equation (\ref{m3m1}) can be conveniently re-casted in the following form
\be
{X_{\star}/Y_{\rm m} \over \sqrt{1+X_{\star}^2/Y_{\rm m}^2}}={m_1\over m_3}\,
{(1-X_{\star})/Y_{\rm m}\over \sqrt{1+(1-X_{\star})^2/Y_{\rm m}^2}},
\ee
from which one can see that
\be
{X_{\star}\over Y_{\rm m}}\leq {m_1\over\sqrt{m_3^2-m_1^2}} \,
\ee
and from this it follows that $X_{\star}\leq m_1/(\sqrt{2}\,m_3)$.
Neglecting terms ${\cal O}(m_1^2/m_3^2)$ one can then
find the approximate relation
\be
X_{\star}\simeq {m_1\over m_3}\, {Y_{\rm m}\over \sqrt{1+Y_{\rm m}^2}},
\ee
while the Eq. (\ref{mtmax}) becomes simply
\be
m_1\,\sqrt{Y_{\rm m}^2+1}+m_3\,Y_{\rm m}=\mt \, .
\ee
It is in the end easy to obtain the expression
\be\label{attempt}
f(m_1,\mt)={m_3-m_1\,\sqrt{1+{m^2_3-m^2_1\over\mt^2}}\over m_3-m_1},
\ee
valid at the first order in $m_1/m_3$, while at the zero-th order in
$m_1/m_3$ and $\mt/m_3$ (thus for $m_{\rm atm}\simeq m_3\gg \mt\geq m_1 $)
it becomes simply $f(m_1,\mt)=1-m_1/\mt$.
The expression (\ref{attempt}) has been  found in \cite{bdp3}
assuming $X_{\star}=0$ and used to get an
upper bound on the absolute neutrino mass scale $m_i\leq 0.1\,{\rm eV}$.
This function clearly does not describe the function
$f(m_1,\mt)$ for any value of $m_1$. In Fig. 5 one can see (dashed lines) how
for increasing values of $m_1/m_{\rm atm}$, from $0.01$ to $1$, the
function saturates to its asymptotic limit $f=1-(m_1/\widetilde{m}_1)^2$
and so, in particular,  it does not reproduce the
correct bound represented by the Eq. (\ref{qd}) \cite{hambye}. However
for values $m_1 \simeq 2\,m_{\rm atm}\simeq 0.1\,{\rm eV}$, those
corresponding to the bound on neutrino masses, it underestimates the bound
on the $C\!P$ asymmetry at most of  $\sim 20\%$ (see Fig. 5 and compare
with the solid line for $m_1/m_{\rm atm}=2$) and
the neutrino mass upper bound itself is therefore underestimated only of $\sim 5\%$
(corresponding to $\sim 0.01\,{\rm eV}$) \cite{bdp4},
much below the precision of the bound itself. This,
together with the result that we have just shown, for which this function
is correct in the limit $(m_1/m_3)^2<1$, fully justifies its approximate use
for an arbitrary value of $m_1$ and is particularly interesting in connection with a
search of the neutrino mass upper bound when some restriction on the space of
relevant parameters is introduced, such that the neutrino mass
bound becomes more stringent and thus it falls in the region $(m_1/m_3)^2<1$.
This is for example the case when a cut-off
$M_1^{\star}\lesssim 10^{12}\,{\rm GeV}$ on the value of $M_1$
is imposed \cite{bdp3,proc}. In Section 6 we will also
show another ueful application of the Eq. (\ref{attempt}).
We have also compared the exact $f(m_1,\mt)$, given by the equations
(\ref{mtmax}), (\ref{m3m1}) and (\ref{fY})
and plotted in Fig. 5 (solid lines), with the results obtained in \cite{hambye}
using the approximation $m_{\rm sol}=0$. We have found a slight
difference, within a few per cent, for $m_1\simeq \mt$,
getting completely negligible as soon as $\mt$ is appreciably
larger than $m_1$.

\section{The effective leptogenesis phase}

In the previous section we have characterized
which are the configurations that maximize the
$C\!P$ asymmetry, or equivalently for which the
effective phase $\sin\delta_L=1$. Here we want to
study how easily this condition can be realized
and which are the typical values of the phase for
different classes of configurations. In general
we can write, from its definition (cf. (\ref{lepphase}),(\ref{bY}),(\ref{f})),
\be\label{genlepphase}
\sin\d_L(m_1,\mt,\O_{j1}^2)={m_1+m_3 \over \mt\,\,f(m_1,\mt)}\,(Y_3+\sigma^2\,Y_2),
\ee
where
$\sigma\equiv\sqrt{\Delta m^2_{21}}/m_{\rm atm}$
is given by $m_{\rm sol}/m_{\rm atm}$ or by $\sqrt{1-m_{\rm sol}^2/m_{\rm atm}^2}$,
depending on whether the case of normal or inverted hierarchy is considered.
The expression (\ref{genlepphase}) can be also recasted as
\be\label{genlepphase2}
\sin\d_L(m_1,\mt,\O_{j1}^2)={m_1+m_3 \over f(m_1,\mt)}\,
\left({\r_3\sin\varphi_3+\sigma^2\,
\r_2\sin\varphi_2 \over m_1\,\r_1+m_2\,\r_2+m_3\,\r_3}\right)\, ,
\ee
showing that the effective phase reduces to
genuine phases in particular cases, while in general it
is given by a linear combination of them, analogously to
how the effective neutrino mass $\mt$ is a linear combination
of the neutrino masses, with coefficients given by
the $\rho_{j}$'s in both cases.
It should be stressed that, in the general framework we are considering,
it does not depend on the phases that could be responsible for $C\!P$
violation at low energies and contained in the mixing matrix $U$.
The requirement of successful leptogenesis implies the lower bound
\be\label{lbsind}
\sin\d_L(m_1,\mt,\O_{j1}^2)
\geq {\eta_{B}^{CMB} \over \eta_B^{\rm max}(M_1,m_1,\mt)} \, ,
\ee
that, together with the lower bounds on $M_1$ and on $T_{\rm reh}$,
represents another  useful way
to get information on the $\O$-parameters through leptogenesis.
For practical purposes it is interesting to study the effective phase
in the two asymptotic limits of {\em fully hierarchical neutrinos}
and {\em quasi-degenerate neutrinos}.

\subsection{Fully hierarchical neutrinos}

In this case one has $m_1=0$, $m_3=m_{\rm atm}$ and $f=1$
\footnote{Notice that then simply one has $\sin\delta_L=\beta$ since $\beta_{\rm max}=1$.}
and the general expression (\ref{genlepphase2}) becomes
\be\label{hierlepphase}
\sin\d_L={m_{\rm atm}\over\mt}\,
\left[\rho_{3}\sin\varphi_3+\sigma^2\rho_2\sin \varphi_2\right] \,,
\ee
where now simply $\mt=m_2\,\rho_{2}+m_{\rm atm}\,\rho_{3}$
and $m_2=m_{\rm sol}\,(\simeq m_{\rm atm})$ in the case of
normal (inverted) hierarchy.
The value of the phase in the full hierarchical case is
relevant in connection with the ($3\,\sigma$) lower bound
on $M_1$ given by
\be\label{lbM1}
M_1 \gtrsim {4.2\times 10^{8}\,{\rm GeV} \over \k_{\rm f}(\mt)\,\sin\d_L}
\hspace{10mm} (M_1\ll 10^{14}\,{\rm GeV})\, ,
\ee
that generalizes the case of maximal phase considered in \cite{bdp1}.
When thermal effects on the Higgs mass are taken into account and
a proper subtraction procedure of on shell $\Delta L=2$ processes is
employed \cite{gnrrs},
the efficiency factor can be approximately
calculated accounting only for decays and
inverse decays \cite{bdp4}. In any case this provides
a conservative upper bound, since other effects,
in particular those arising from spectator processes \cite{spectator},
can only lower the final asymmetry.  Then
a very good analytical approximation in
the case of an initial thermal abundance, that gives quite a
conservative upper limit in the weak wash-out regime, is
\be\label{kf}
k_{\rm f}(\mt)\simeq {2\over K\,z_B(K)}\,\left(1-e^{-{K\,z_B(K)\over 2}}\right)  \, ,
\ee
where the {\em decay parameter} $K$ is related to the  effective neutrino mass through the relation
\be
K={\mt\over m_{\star}}\simeq 926\,{\mt\over {\rm eV}} \, ,
\ee
while the {\em baryogenesis $z\equiv M_1/T$ value} is calculated using
\footnote{This is a different
expression from the two ones, practically equivalent, given
in  \cite{bdp4} and in \cite{proc} and is such to cure
the error of the expression (\ref{kf}) around values $K\sim 1$,
yielding an analytical expression that well agrees with the numerical
calculation within $10\%$ for all values of $\mt$.}
\be\label{zB}
z_{B}(K)\simeq 9/8+2.35\,[\ln(1+K)]^{0.8} \, .
\ee
A comparison between the analytical approximation
(cf. (\ref{kf}) and (\ref{zB})) and the numerical solution is shown in Fig. 7
while the lower bound on $M_1$ is plotted in Fig. 8.
\begin{figure}[t]
\centerline{\psfig{file=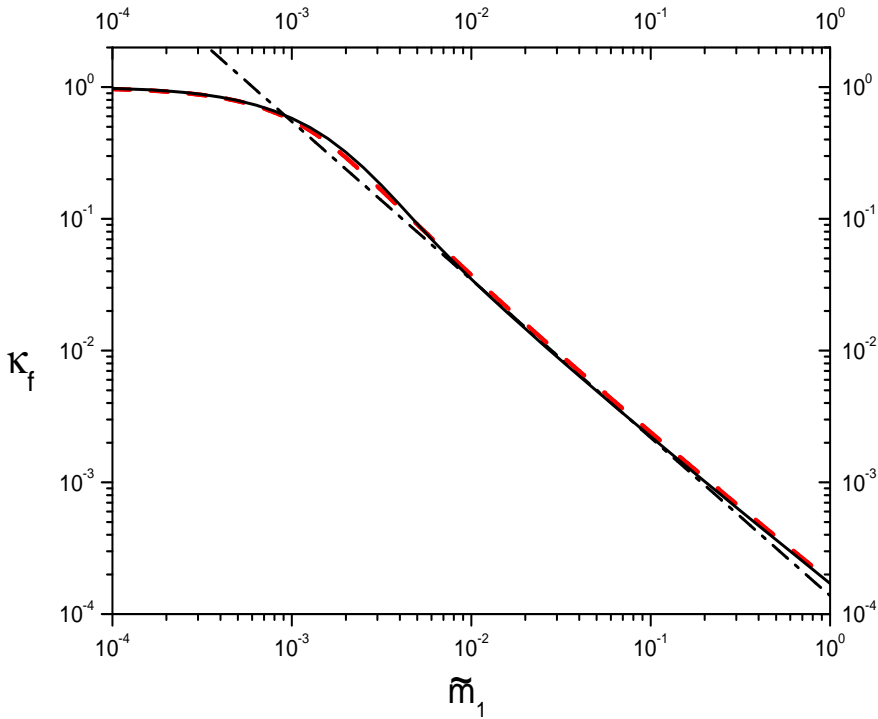,height=10cm,width=14cm}}
\vspace{-17mm} \caption{\small Efficiency factor: the analytic approximation
(dashed line) given by the eq.'s (\ref{kf}) and (\ref{zB}) is compared with the
numerical solution (solid line) of the kinetic equations when the wash-out
from scatterings is neglected and an initial thermal $N_1$ abundance
is assumed \cite{bdp4}. The dot-dashed line is the power-law fit
eq. (\ref{kfsw}).}
\end{figure}

\begin{figure}[t]
\centerline{\psfig{file=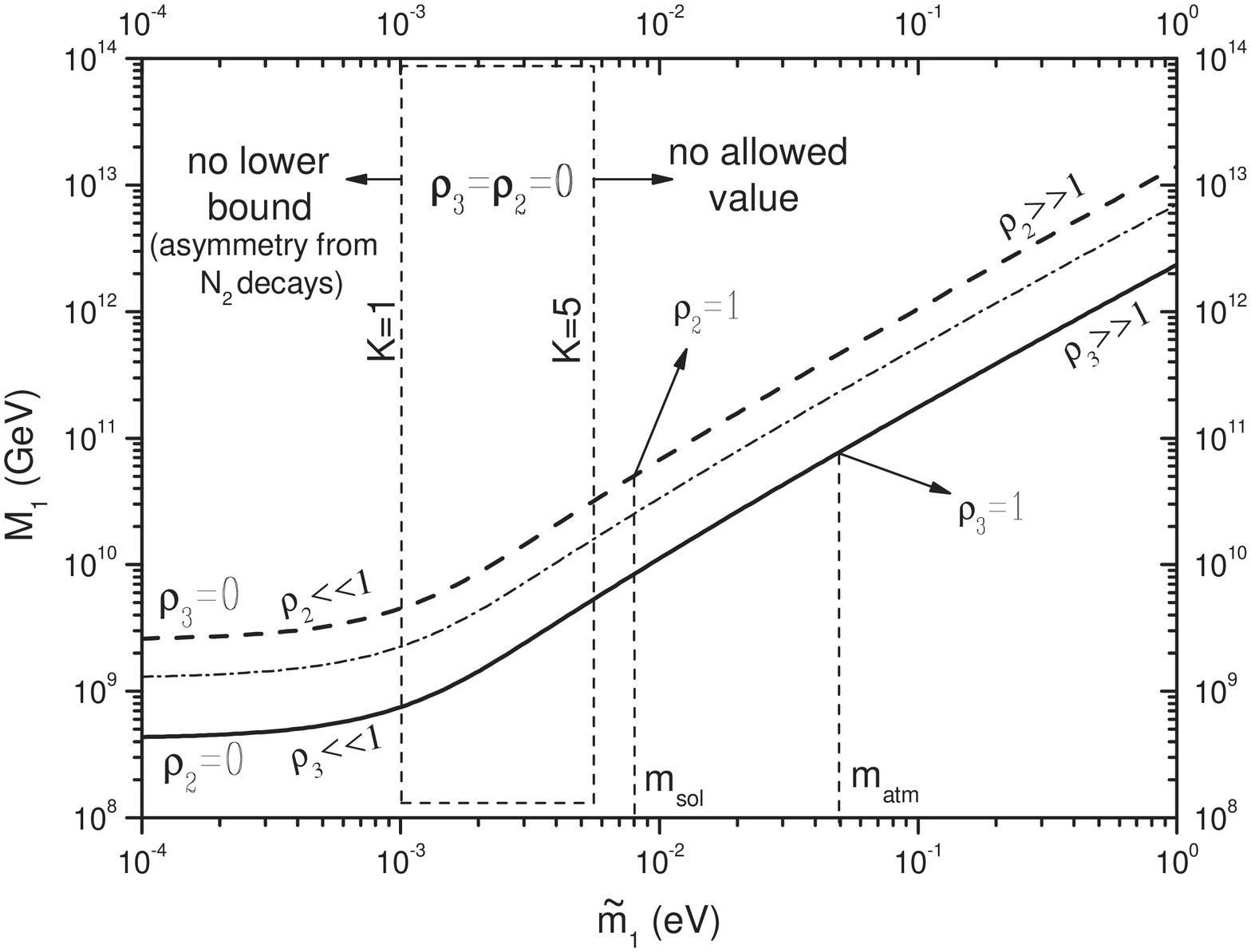,height=115mm,width=14cm}}
\vspace{-11mm} \caption{\small Lower bounds ($3\sigma$) on $M_1$
for different value of the effective phase in the case of normal
hierarchy. The solid line corresponds to maximal phase in the case
$\rho_2=0$ (cf. (\ref{lbM1})). The dashed line corresponds to the
case $\rho_3=0$ and is given by the Eq. (\ref{lbM1}) with
$\sin\d_L=\sigma\simeq 1/6$. The dot-dashed line is the
intermediate case $\rho_3=\sigma^2\,\rho_2$ with an effective
phase $\sin\delta_L=2\,\sigma\simeq 1/3$. The box indicates a
transition range of $\mt$ values in the scenario where the
asymmetry is generated by the $N_2$'s decays for $\r_3=\r_2=0$:
for $\mt\lesssim m_{\star} \,(K\lesssim 1)$ there is no lower bound on $M_1$,
while for $\mt\gtrsim 5\,m_{\star} \,(K\gtrsim 5)$  there is no allowed value.}
\end{figure}
For $0.5\,m_{\rm sol}\lesssim \mt \lesssim 2\,m_{\rm atm}$
the expression (\ref{kf}) can be approximated with the power law \cite{proc}
\be\label{kfsw}
k_{\rm f}(\mt) \simeq {0.5 \over K^{1.2}}\simeq
3.5\times 10^{-2}\,\left( {10^{-2}\,{\rm eV}\over \mt} \right)^{1.2} \,
\ee
and in the range $\mt=[m_{\rm sol},m_{\rm atm}]$ one gets,
from the Eq. (\ref{lbM1}), $M_1\gtrsim (10^{10}-10^{11})\,{\rm GeV}$.
Notice that only for $K\gtrsim 5$ the lower bound on $M_1$
does not depend on the initial conditions and in particular
on a specific description of $N_1$'s production. The lower bound
at $K\simeq 5$, $M_1\gtrsim 5\times 10^{9}\,{\rm GeV}$,
is thus particularly meaningful, since it
can be regarded as the lowest allowed $M_1$ value without the
need of additional assumptions beyond the given ones.

Within the {\em minimal supersymmetric standard model} (MSSM) the
$C\!P$ asymmetry and its maximum value get double and this implies
a relaxation of the lower bound Eq. (\ref{lbM1}) at low $\mt$,
such that \cite{proc}
\be
M_1 \gtrsim {2.25\times 10^{8}\,{\rm GeV}\over \k_{\rm f}(\mt)\,\sin\d_L}
\hspace{10mm} (M_1\ll 10^{14}\,{\rm GeV})\,.
\ee
The expression (\ref{kf})  for the efficiency factor is still valid
but the stronger wash-out processes make
the equilibrium neutrino mass a  bit lower than in the
SM case, $m_{\star}^{MSSM}\simeq 0.0008\,{\rm eV}$, such that
$K\simeq 1270\,\mt/{\rm eV}$. In this way
the efficiency factor in the strong wash out regime gets also lower
and the power law (\ref{kfsw}) becomes \cite{proc}
\be
k_{\rm f} \simeq {0.5\over {K}^{1.2}}
\simeq 2.4\times 10^{-2}\,\left( {10^{-2}\,{\rm eV}\over \mt} \right)^{1.2} \, ,
\ee
while the $3\sigma$ lower bound on $M_1$ in the range $\mt=[m_{\rm sol},m_{\rm atm}]$
becomes just slightly more relaxed. Within a thermal picture
of leptogenesis, as we are assuming, the lower bound on $M_1$ is
closely related to a lower bound on the reheating temperature
of the early Universe $T_{\rm reh}$ that is approximately
given by the expression \cite{bdp4}
\be\label{lbTreh}
T_{\rm reh}^{\rm min}\simeq {M_1^{\rm min}\over z_B-2\,e^{-\alpha/K}} \, ,
\ee
plotted in Fig. 9 for $\alpha=0.5$
\footnote{The result is more conservative compared both to the
numerical results presented \cite{gnrrs} and to the analytical
ones presented in \cite{proc}:
the wash-out from scatterings is completely neglected,
the approximation (\ref{kf}) slightly overestimates (within $10\%$)
the numerical results and the approximation (\ref{lbTreh})
underestimates $T_{\rm reh}$ at $K\lesssim 4$. In this way the
restrictive bounds we obtain for models with phase suppression,
can be regarded as very conservative ones. Note also that we are using a different
value of $\alpha$ compared to \cite{bdp4} as a consequence
of the different expression employed for $z_B(K)$.}.
\begin{figure}[t]
\centerline{\psfig{file=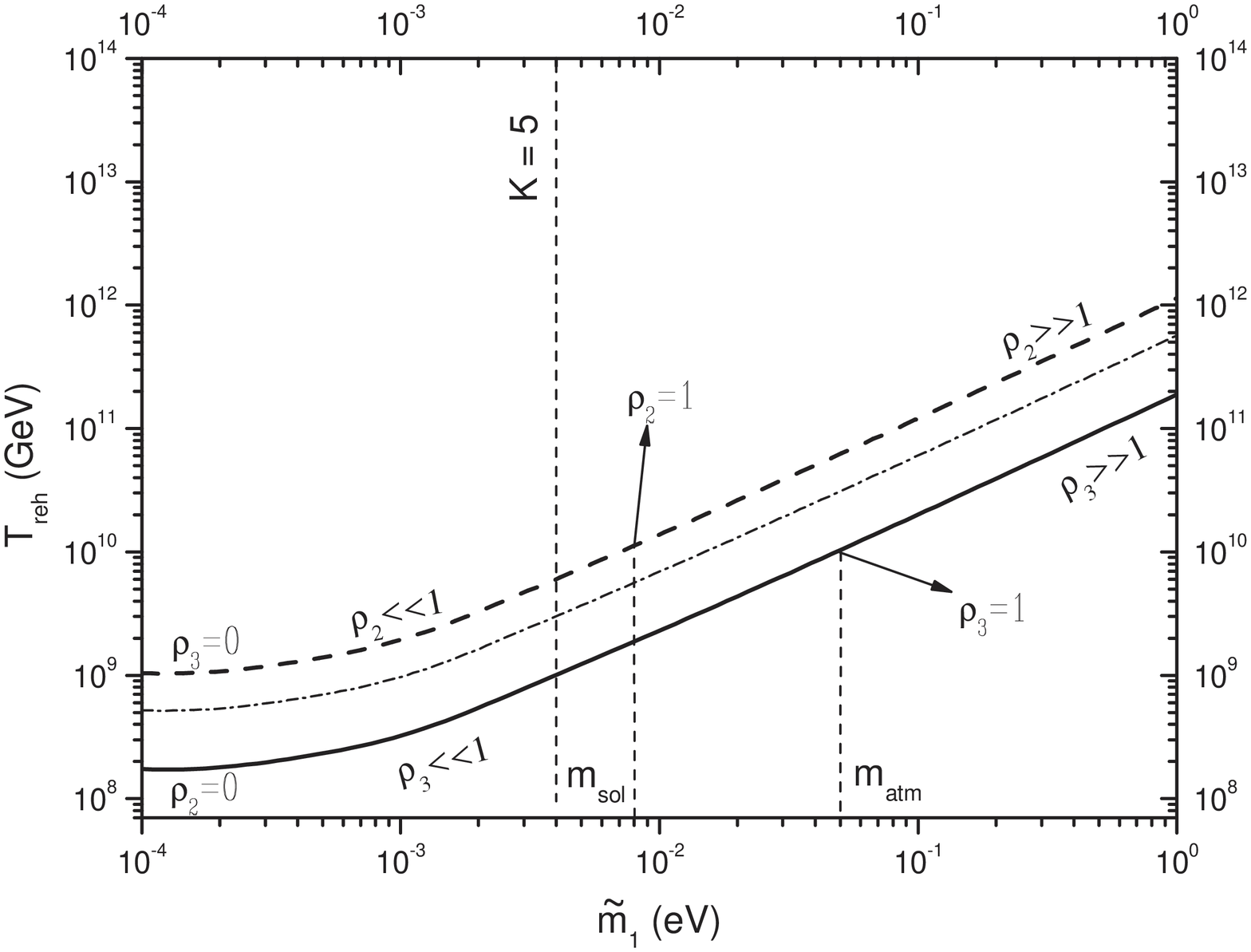,height=12cm,width=14cm}}
\vspace{-10mm}
\caption{\small Lower bound ($3\,\sigma$)
on $T_{\rm reh}$ in the MSSM case (cf. (\ref{lbTreh})).
The lines correspond to the same cases of fig 7. For values
 $K\gtrsim 5$ (right of the dashed line) the lower bound
does not depend on the initial conditions.}
\end{figure}
For $K\simeq 5$ one has the lowest $T_{\rm reh}$ value
independent on the choice of initial conditions and conservatively
we obtain $T_{\rm reh}\gtrsim 10^{9}\,{\rm GeV}$.

The existence of such a lower bound on $T_{\rm reh}$ represents a problem
within the MSSM picture, since this has to be compared with the upper bound
from the avoidance of the gravitino problem. Values of the gravitino mass
$\ll 100\,{\rm TeV}$ require reheating temperature
$T_{\rm reh}\ll 10^{10}\,{\rm GeV}$  \cite{gravitino1}.
From Fig. 9 one can see that there is compatibility only for
values $\mt\ll m_{\rm atm}$.

These results are valid under the conservative assumption of maximal phase.
In general, if one modifies the picture such that
the final predicted asymmetry $\eta_{B}\rightarrow \xi\,\eta_B$, then
the lower bound on $M_1$ and on the reheating temperature change as
$(M_1^{\rm min},T_{\rm reh}^{\rm min})\rightarrow (M_1^{\rm min},T_{\rm reh}^{\rm min})/\xi$
\cite{bdp2}. Therefore, if we consider models with some phase suppression, $\sin\delta_L<1$,
the minimum allowed values of $M_1$ and $T_{\rm reh}$ increase
compared to the case of maximal phase. On the other hand, if
$M_1^{\rm min}$ becomes larger than about $10^{14}\,{\rm GeV}$,
the assumption that $\k_{\rm f}$
depends only on $\mt$ is not valid any more, since $\Delta L=2$ processes
give an additional contribution to the wash-out depending also on $M_1$ \cite{bdp1}.
In this case the efficiency factor is exponentially suppressed at large $M_1$
values such that
\be
\k_{\rm f}(\mt)\rightarrow \k_{\rm f}(\mt)\,e^{-{\o\over z_B}\,{M_1\over 10^{14}\,{\rm GeV}}} \, ,
\hspace{10mm} (\o\simeq 5)
\ee
and $\eta_B^{\rm max}(M_1,m_1=0,\mt)$ (cf. (\ref{etaBmax})) has a maximum at
$M_1\simeq 2\,z_B\,10^{13}\,{\rm GeV}$ given by \cite{aspects}
\be
\eta_B^{\rm max}(m_1=0,\mt)\simeq
1.5 \times 10^{-5}\,{10^{-3}\,{\rm eV}\over \mt} \,\,\;\;\;\,\,\,\,\,
(\mt\gtrsim 10^{-3}\,{\rm eV}) \, .
\ee
This implies that the lower bound on the
phase (cf. (\ref{lbsind})) cannot be arbitrarily evaded increasing $M_1$ and
one can place an interesting lower bound, independent
on the value of  $M_1$, given by
\be\label{phaselb}
\sin\d_L \geq \, 4\times 10^{-5}\,{\mt\over 10^{-3}\,{\rm eV}} \,
\hspace{10mm} \, (\mt\gtrsim 10^{-3}\,{\rm eV}) \, .
\ee
Let us study {\em two special cases}.
For $\rho_{2}=0$ one has $\mt= \r_3\,m_{\rm atm}$ and from the Eq. (\ref{hierlepphase})
\be
\sin\delta_L=\sin\varphi_3 \, .
\ee
One can set $\delta_L=\varphi_{3}$, fixing
unambiguously $\delta_L$. Again one can see that
the effective leptogenesis phase is maximal for $\rho_{2}=0$ (i.e. $\O_{21}=0$)
and $X_3=0$, corresponding to $\varphi_{3}=\pi/2$.
For random assignations of $\varphi_3$, the average value of the phase,
$\langle|\sin\varphi_3|\rangle=2/\pi \simeq 0.6$,
suggests that a value close to the maximal one is a reasonable possibility
not requiring any particular tuning.

Within models with single or sequential RH neutrino dominance,
characterized by $|X_j|\simeq 1 \gg |X_{i\neq j}|$,
there are two choices: either $X_1\simeq 1\gg |X_3|$
or conversely $X_3\simeq 1\gg |X_1|$. The first one
corresponds to maximal phase $\sin\delta_L\simeq 1$, while
$\mt\simeq Y_3\,m_{\rm atm}$. If furthermore one has $Y_3\ll 1$,
then one has the kind of models we previously discussed, with $\O$ obtained as
a small complex perturbation of the matrix (\ref{O23}) and
small values of $\mt\ll m_{\rm atm}$,
such that minimum values $M_1,T_{\rm reh}\sim 10^{9}\,{\rm GeV}$ are possible.
In the second case, i.e. $X_3\simeq 1\gg |X_1|$, such that $N_1$ contributes
only to $m_3$, one has necessarily some phase
suppression and moreover $\mt\geq m_{\rm atm}$.
The value of the phase can be expressed as a function of
$\mt$ and is given by $\sin\d_L=\sqrt{1-{m_{\rm atm}^2/\mt^2}}$.
Since the efficiency factor is approximately  inversely
proportional to $\mt$ (cf. (\ref{kfsw})), then one easily finds that the final asymmetry
$\eta_B(m_1=0,\mt)\propto \k_{\rm f}(\mt)\,\sin\d_L(\mt)$ is maximized
for $\mt\simeq \sqrt{2}\,m_{\rm atm}$ corresponding to $\sin\d_L\simeq 1/\sqrt{2}$.
This implies a much more stringent lower bounds on $M_1$ and on $T_{\rm reh}$
that can be read from Fig.'s 8 and 9 obtaining
$M_1\gtrsim 1.5\times 10^{11}\,{\rm GeV}$ and $T_{\rm reh}\gtrsim 2\times 10^{10}\,{\rm GeV}$.
Moreover the lower bound on the phase (cf. (\ref{phaselb})) becomes
$\sin\phi_3\gtrsim 2\times 10^{-3}$.

Still another possibility is that $N_1$ contributes both
to $m_1$ and to $m_3$ and in this case $|X_1|\sim |X_3|\gtrsim 1$,
such that $\mt\gtrsim m_{\rm atm}$, still
implying $M_1\gtrsim 10^{11}\,{\rm GeV}$ and $T_{\rm reh}\gtrsim 10^{10}\,{\rm GeV}$.

The second special case is for $\rho_{3}=0$, such that $\mt=m_2\,\rho_2$
and from the Eq. (\ref{hierlepphase}) one obtains
\be\label{lpr30}
\sin\d_L=\sigma\sin\varphi_2 \, .
\ee
If one considers the case of {\em normal hierarchy}, then
$\sigma=m_{\rm sol}/m_{\rm atm}\simeq 1/6$ and this
unavoidably leads to a strong phase suppression,
$\sin\delta_L\simeq \sin\varphi_2/6 \ll 1$,
that results in a
more stringent bound on $M_1$ (see the dashed line in Fig. 8)
and on $T_{\rm reh}$ (see the dashed line in Fig. 9).
For models with $N_1$ contributing only to $m_2$,
such that  $X_2\simeq 1\gg |X_1|$, the
phase suppression is $\sin\d_L < 1/6$,
and similar lower bounds for $M_1$ and $T_{\rm reh}$, as in
the case of $N_1$ contributing only to $m_3$, hold, even though
the wash-out is much weaker because now
$\mt\simeq m_{\rm sol}\ll m_{\rm atm}$.

 We have thus two extreme cases, $\rho_2=0$ or $\rho_3=0$,
where the effective phase can be maximal or remarkably suppressed.
Let us now analyze the general case, with both non-vanishing $\rho_2$ and $\rho_3$.
Since we are interested in the lower bound on $M_1$,
we can study the expression  (\ref{hierlepphase}) for maximal phases,
$\sin\varphi_2=\sin\varphi_3=1$, and thus
\be
\sin\delta_L={\rho_3+\sigma^2\rho_2\over \rho_3+\sigma\rho_2} \,.
\ee
These are the most general models with
$N_1$ contributing only to $m_1$, such that
$|X_2|,|X_3|\ll X_1\simeq 1$.
For $\rho_3 \gg \sigma\rho_2$ one has $\sin\delta_L\simeq 1$ and
$\mt\simeq \rho_3\,m_{\rm atm}$, such that the limit $\rho_2=0$ is recovered, while
conversely, for $\rho_3\ll \sigma^2\,\rho_2$, one has the
phase suppression $\sin\delta_L=\sigma\simeq 1/6$
and $\mt\simeq \rho_2\,m_{\rm sol}$ and one recovers the limit $\rho_3=0$.

In the intermediate case, $\sigma^2\rho_2\lesssim \rho_3 \lesssim \sigma\rho_2$,
one has $2\,\sigma\lesssim \sin\d_L\lesssim 1/2$.
For example for $\rho_3=\sigma\rho_2$ one has $\mt=2\,m_{\rm sol}\simeq 0.016\,{\rm eV}$
and $\sin\delta_L=1/2$ corresponding to $M_1 \gtrsim 4\times 10^{10}\,{\rm GeV}$
and $T_{\rm reh} \gtrsim 7\times 10^{9}\,{\rm GeV}$. For
$\rho_3=\sigma^2\,\rho_2\ll 1$ one has $\sin\delta_L=2\,\sigma\simeq 1/3$
and $\mt\simeq m_{\rm sol}\,\rho_2$. The corresponding lower bounds
on $M_1$ and $T_{\rm reh}$ are shown in Fig. 8 and 9 (dot-dashed lines).
If $\rho_2=1$ one gets
$M_1\gtrsim 3\times 10^{10}\,{\rm GeV}$ and
$T_{\rm reh}\gtrsim 5\times 10^{9}\,{\rm GeV}$.
Therefore, this case represents the best compromise to get the lowest possible values for $T_{\rm reh}$
and $M_1$ without having $\rho_2,\rho_3\ll 1$ and $\rho_1\simeq 1$.

We can conclude that {\em the possibility to access the window
$10^{9} \lesssim M_1\,(T_{\rm reh})/{\rm GeV}  \lesssim (0.5)\,3\,\times \, 10^{10}$
relies on a particular class of models}, such that $\r_2,\r_3\ll \r_1 \simeq 1$,
the only ones  allowing both maximal phase and $\mt\ll m_{\rm atm}$.
This conclusion does not change
if one considers the case of {\em inverted hierarchy}:
it is true that one does not have
phase suppression but values of $\mt\ll 10^{-2}\,{\rm eV}$
are obtained again only for $\rho_2,\rho_3 \ll 1\simeq \rho_1$.

More generally values $T_{\rm reh}\lesssim 10^{10}\,{\rm GeV}$ are obtained
only within models where $N_1$ contributes only to $m_1$. Among sequential
dominated models these correspond to models where $m_1$ is dominantly
determined just by $N_1$.
For models where $N_1$ contributes only to $m_3$, including
sequential dominated models with $m_3$ dominated by $N_1$,
we have seen that, if $\rho_2=0$, then necessarily the lower bounds
are more restrictive, such that $M_1\gtrsim 1.5\times 10^{11}\,{\rm GeV}$
and $T_{\rm reh}\gtrsim 2\times 10^{10}\,{\rm GeV}$. Allowing
a non zero $Y_2$ the bounds can only become more restrictive.
On the other hand if one considers models where $N_1$ contributes
dominantly only to $m_2$ ($X_2\simeq 1\gg X_1,X_3$),
including those where $m_2$ is dominated by $N_1$,
the limits obtained for $\r_3=0$
can be relaxed if one allows a non zero $Y_3$. The best
optimization between high phase and low $\mt$ is obtained for
configurations with $Y_2=0$. In this case one has $\sin\d_L=1-m_{\rm sol}/\mt$
and it is simply to see that the asymmetry is maximized
for $\mt\simeq 2\,m_{\rm sol}$ corresponding to $\sin\d_L\simeq 0.5$.
One then obtains $M_1\gtrsim 4\times 10^{10}\,{\rm GeV}$ (within SM) and
$T_{\rm reh}\gtrsim 7\times 10^{9}\,{\rm GeV}$ (within MSSM)
\footnote{An interesting exercise is also to consider the limit $M_3\rightarrow\infty$,
as already studied in \cite{ct}. In this situation the $\O$ matrix is simply given by
\be
\O=
\left(
\begin{array}{ccc}
     0     & 0 &  1 \\
\O_{21}  & - \sqrt{1-\O^2_{21}} & 0 \\
\sqrt{1-\O^2_{21}} &  \O_{21} & 0
\end{array}
\right)
\, .
\ee
This implies that $Y_2=-Y_3$ and from the eq. (\ref{hierlepphase}) one can see
that for inverted hierarchy there is a large phase suppression.
For normal hierarchy, neglecting ${\cal O}(\sigma^2)$ terms, one has
that $\sin\d_L\simeq m_{\rm atm}\,Y_3/\mt$. The maximum of $Y_3$
can be found analogously to the general case simply replacing
$m_1$ with $m_2$ in the eq. (\ref{fY}), such that
$Y_3^m=f(m_2,\mt)/(m_{\rm atm}\,(1+\sigma))$ (where now
$\mt\geq m_2$).
It is then simple to find that $\eta_B$ is maximized for $\mt\simeq 2\,m_{\rm sol}$
and that $\sin\d_L\simeq 0.4$, such that
$M_1\gtrsim 5\times 10^{10}\,{\rm GeV}$ (within the SM)
and $T_{\rm reh}\gtrsim 9\times 10^{9}\,{\rm GeV}$ (within the MSSM).
These results are more relaxed than in \cite{ct}, simply because we
are taking into account the recent results of \cite{gnrrs,bdp4}
on the efficiency factor and on the
relation between $M_1^{\rm min}$ and $T_{\rm reh}^{\rm min}$,
as previously discussed.}.


\subsection{Quasi-degenerate neutrinos}

In this case one has $m_1\simeq m_2 \simeq m_3 $ and thus $m_1/\mt \simeq {\ell}^{-1}$,
where remember that ${\ell}\equiv \rho_1+\rho_2+\rho_3\geq 1$.
The expression (\ref{qd}) can then be re-casted as
$f\simeq \sqrt{{\ell}^2-1}/{\ell}$ and the Eq. (\ref{genlepphase2}) for the phase becomes
\be\label{phasedeg}
\sin\delta_L\simeq {2\over\sqrt{\ell^2-1}}\,
[\rho_3\sin\varphi_3+\sigma^2\rho_2\sin\varphi_2] \, .
\ee
For $\rho_2=0$ and $X_3=1/2$ one has
$Y_3=2/\sqrt{\ell^2-1}$ and the case of
maximal phase is recovered. On the other hand if one considers models
with $\rho_3=0$, then the maximum asymmetry is still
realized for $X_2=1/2$  but this time one has $\sin\delta_L=\sigma^2$,
meaning that if $\sigma=m_{\rm atm}/m_{\rm sol}\simeq 1/6$
(`normal' quasi-degenerate spectrum),  then one has a strong  phase suppression,
$\sigma$  times stronger than in the case of fully hierarchical neutrinos.

We can also consider an intermediate case with both $\rho_2$
and $\rho_3\neq 0$, for example when
$\varphi_2=\varphi_3$ and $X=1/2$ (dot-dashed arrows in Fig. 10).
\begin{figure}[t]
\centerline{\psfig{file=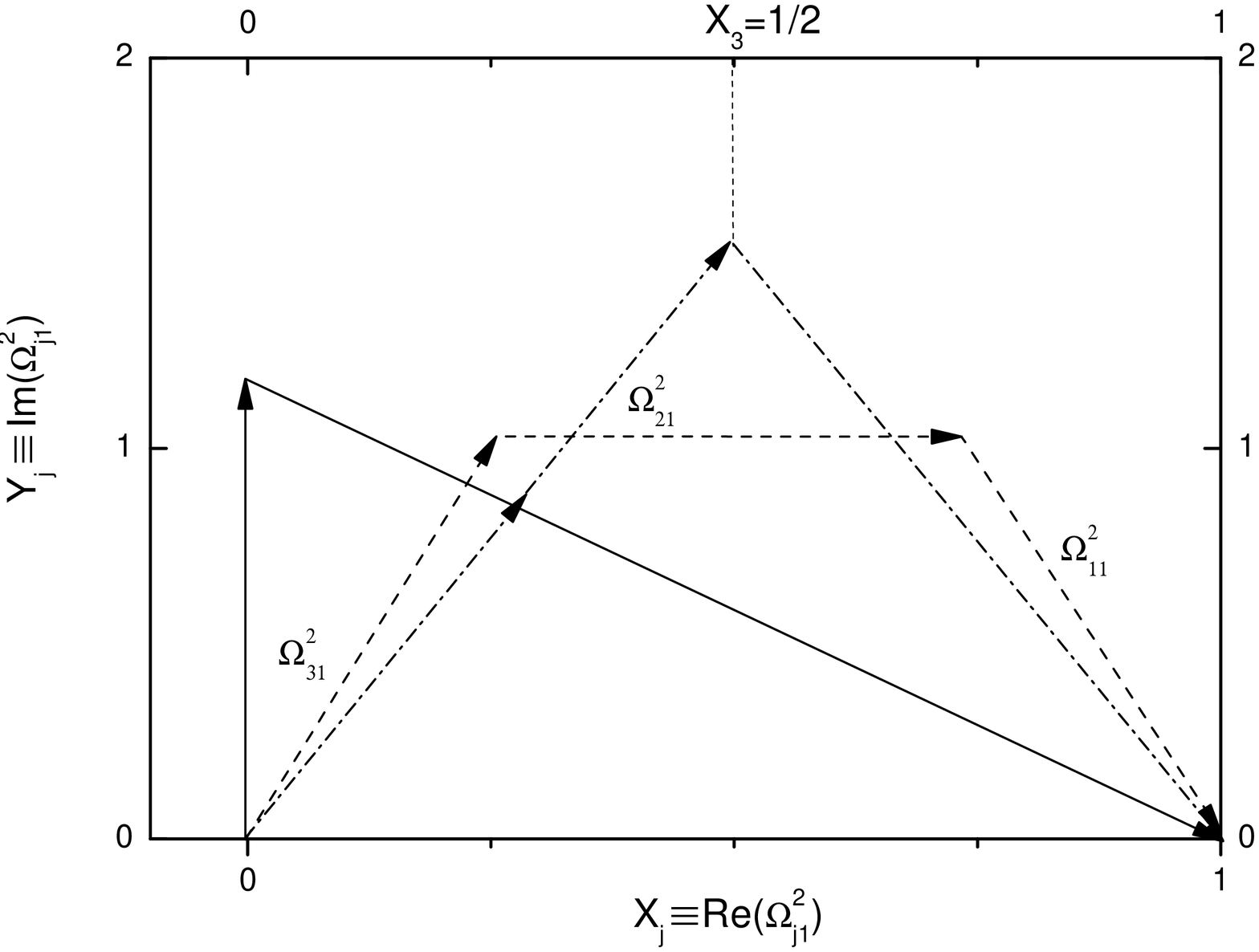,height=13cm,width=12cm}}
\vspace{-15mm}
\caption{\small quasi-degenerate neutrinos; three different cases
with non maximal phase, as considered in the main text.}
\end{figure}
In this case the general expression (\ref{phasedeg}) gets specialized into
$\sin\d_L=1-\zeta+\sigma^2\,\zeta$,
where $\zeta=\rho_2/(\rho_2+\rho_3)$, and describes the transition between
the case with $\rho_2=0$ and the case with $\rho_3=0$.

Another example is given by a situation where $\rho_2=0$
but the condition $X_3=1/2$ is not realized. For example if one takes
$X_3=0$ (see solid arrows in Fig. 10), then one gets $\sin\d_L=\sqrt{1-1/{\ell}^2}$.
For ${\ell}\gg 1$, corresponding to $\mt\gg m_1$, one recovers maximal phase,
while for finite values of $\mt$ there is some phase suppression.
An interesting example is to take $\mt=\sqrt{2}\,m_1$,
since this corresponds to the peak value that saturates the upper bound
on neutrino masses \cite{bdp4}. In this particular case one finds
$\sin\d_L=\sqrt{1/2}\simeq 0.7$. An analogous result holds if one takes
$X_1=0$ and $X_3=1$. Notice that these two cases correspond to a lightest neutrino
mass dominated by the lightest RH neutrino (for $X_3=0$) or by the
heaviest (for $X_1=0$).

We have also considered the case of configurations with $Y_2=0$
and $X_3=(1-X_2)/2$, such that when $X_2=0$ one recovers the
configuration with maximal phase (dashed arrows in Fig. 10).
In this case one obtains easily
\be
\sin\d_L=\sqrt{1-{2\,X_2\over \ell + 1}} \,
\ee
and for example, for $\rho_2={\ell}/3$, one gets $\sin\d_L\simeq 0.7$. These
examples show that typically a phase suppression is expected. This is relevant
in connection with the neutrino mass upper bound, $m_i\leq 0.1\,{\rm eV}$,
because it implies that the saturation of the bound actually occurs, not only
for special values of $M_1$ and $\mt$, but also for special $\O_{j1}$'s configurations.

\section{A new scenario of thermal leptogenesis} 

So far we have assumed that the final asymmetry is not
influenced by the two heavier RH neutrinos and this
has considerably reduced the number of seesaw parameters
on which the final asymmetry depends. This assumption holds
if the asymmetry generated by the $N_2$'s and by the $N_3$'s is either negligible or is
efficiently washed-out by the inverse decays of the $N_1$'s and also if
the approximated expression (\ref{eps1h}) for the $C\!P$ asymmetry $\ve_1$
can be used. As we will discuss in Section 7, these
two conditions require some degree of hierarchy in the
heavy neutrino spectrum.
Here, even though we still assume a hierarchical heavy neutrino spectrum,
we study a scenario where the final asymmetry is
produced by the $N_2$'s decays. This is possible if the
value of the effective neutrino mass lies  in the weak
wash-out regime. In this case
the wash-out from $N_1$ inverse decays and scatterings is not sufficient
to wash-out a previously generated asymmetry, in particular that one
generated from the $N_2$'s decays.
As we discussed in Section 2, the possibility
that $\mt\lesssim m_{\star}\simeq 10^{-3}\,{\rm eV}$,
relies on models with an $\O$ matrix obtained as a
small perturbation of a 23-rotation in the complex plane (cf.(\ref{O23})).
Naively one could think that, since we
are still assuming a hierarchical spectrum of neutrinos such that $M_2\gg M_1$,
and because of the lower bound on $M_1$, one has to require large
values of the reheating temperature such that $T_{\rm reh}\gtrsim M_2\gg 10^{9}\,{\rm GeV}$.
However, it is easy to understand that actually the lower bound on $M_1$
disappears and is simply replaced by an analogous lower bound on $M_2$.
First of all let us notice that in a general case the final asymmetry
is the sum of the contributions from all the three $N_i$'s and, assuming a
sufficiently hierarchical heavy neutrino spectrum, it can be written as
\be\label{asytot}
\eta_{B}={a_{\rm sph}\over  N_{\gamma}^{\rm rec}}\; \sum_i\,\ve_i\;\k_{\rm f,i}  \, ,
\ee
that generalizes the expression (\ref{etaB}) with $\k_{\rm f,1}\equiv \k_{\rm f}$.
Let us focus on the asymmetry produced by the $N_2$ decays.
These will be described by a quantity, $\mtt$, that plays
an analogous role of $\mt$ in the description of the
$N_1$ decays and is defined as
\be
\widetilde{m}_2 \equiv {\left( m^{\dagger}_D\,m_D \right)_{22}\over M_2}=
\sum_i\,m_i\,\rho_{i2} \, .
\ee
For our purposes the $\O$ matrix can be conveniently
parameterized in the following way
\be\label{O}
\O=
\left(
\begin{array}{ccc}
 \sqrt{1-\O_{21}^2-\O_{31}^2} & \O_{12} &  - \sqrt{\O_{21}^2+\O_{31}^2-\O^2_{12}} \\
 \O_{21} & \O_{22} & - \sqrt{1-\O_{22}^2-\O_{21}^2} \\
 \O_{31} & \sqrt{1-\O_{22}^2-\O_{12}^2} & \sqrt{\O_{22}^2+\O^2_{12}-\O_{31}^2)}
\end{array}
\right) \, .
\ee
The orthogonality also implies that each product of
different columns or rows has to vanish. Therefore, we can still express one
of the four left complex elements as a function of the other three.
For example, if $\O_{21}\neq 1$, one can write
\footnote{
Alternatively, if $\O_{21}=1$ but $\O_{31}\neq 0$, one has
\be
\O_{12}={\O^2_{31}(1-\O^2_{22})-\O^2_{22}\over 2\,i\,\O_{22}\,\O_{31}} \, .
\ee
In the particular case that $\O_{21}=1$ and $\O_{31}=0$, then
$\O_{22}=0$ and the $\O$ matrix specializes into
\be
\O=
\left(
\begin{array}{ccc}
 0 & \O_{21} &  -\sqrt{1-\O^2_{21}} \\
    1 & 0 & 0 \\
    0 & \sqrt{1-\O^2_{21}} & \O_{21}
\end{array}
\right) \, .
\ee}
\be\label{O12}
\O_{12}= {-\O_{21}\,\O_{22}\,\sqrt{1-\O_{21}^2-\O_{31}^2} \pm
\O_{31}\sqrt{1-\O_{21}^2-\O_{22}^2}\,\over 1-\O_{21}^2}  \, .
\ee
Since we are assuming $\rho_1\simeq 1$ and $\rho_2,\rho_3\ll 1$,
this expression simplifies into
\be
\O_{12}\simeq -\O_{21}\,\O_{22}\pm \O_{31}\,\sqrt{1-\O_{22}^2} \, ,
\ee
from which one can deduce that $\rho_{12}\leq (\rho_{21}+\rho_{31})\,(1+\rho_{22})$,
implying that it is impossible to have $\rho_{12}\simeq 1$
and $\rho_{22}\ll 1$. Therefore, if $\mt\ll m_{\star}$, from the orthogonality
of $\O$ it follows that is impossible to have $\mtt \ll m_{\star}$ too.
 This means that if the asymmetry from the $N_1$'s is generated in the
weak wash-out regime, then the asymmetry from the $N_2$'s
has necessarily to be generated in the strong wash-out regime.
This is an interesting result because, if the final asymmetry is generated
by the $N_2$'s, then it will be independent
on the initial conditions exactly like in the typical case
when the final asymmetry is generated by the $N_1$'s and $\mt$ lies in the
strong wash-out regime
\footnote{This result is also interesting because it suggests
that after all an initial asymmetry is always washed out even if the
lightest RH neutrino decays occur in the weak wash-out regime and so that
from this point of view there is no problem with the initial conditions.
However, in this case one has to impose that the reheating temperature is higher
than $\sim M_2/z_B(K_2)$ and so the lower bound on $T_{\rm reh}$ would be
much more restrictive. Moreover the problem of the dependence on the initial
$N_1$ abundance in the weak wash-out regime still remains.}.

Moreover notice
that a possible asymmetry generated by the heaviest RH neutrinos,
the $N_3$'s, is always washed-out and negligible within the assumption of
a hierarchical RH neutrino spectrum.

We have now to calculate the final asymmetry generated by the
$N_2$'s and see whether this can explain the observed baryon asymmetry.
Therefore, we have to calculate the $C\!P$ asymmetry associated to the
$N_2$ decays. The expression for $\ve_2$ is simply given by an expression
similar to the Eq. (\ref{eps1g}),
\be\label{eps2g}
\varepsilon_2 \simeq  -{1\over 8\pi}
\sum_{i=1,3}\,{{\rm Im}\,
\left[(h^{\dagger}\,h)^2_{2i}\right]\over (h^{\dagger}\,h)_{22}} \,\times\,
\left[f_V\left({M^2_i\over M^2_2}\right)+f_S\left({M^2_i\over M^2_2}\right)\right] \, .
\ee
This time however it cannot be written in the form (\ref{eps1h}),
since now the two terms have to be evaluated in two different limits:
the first one ($i=1$) for $M_1^2/M_2^2 \ll 1$  such that
\be
f_V\left({M_1^2\over M_2^2}\right)+f_S\left({M_1^2\over M_2^2}\right)
\simeq -\,{M_1\over M_2}\,\left(\ln{M_2\over M_1}-2\right)
\ee
and the second one ($i=3$) in the limit $M_3^2/M_2^2 \gg 1$
such that
\be
f_V\left({M_3^2\over M_2^2}\right)+f_S\left({M_3^2\over M_2^2}\right)
\simeq -{3\over 2}\,{M_2\over M_3} \,.
\ee
In these limits the Eq. (\ref{eps2g}) becomes
\be\label{eps2h}
\ve_2\simeq {1\over 8\pi\,(h^{\dagger}\,h)_{22}}
\left[{3\,M_2\over 2\,M_3}\,{\rm Im}[(h^{\dagger}h)^2_{23}]+{M_1\over M_2}\,
\left(\ln{M_2\over M_1}-2\right)\,{\rm Im}[(h^{\dagger}\,h)_{21}^2]\right] \, .
\ee
From the Eq. (\ref{Omega}) one can easily derive the relation
\be\label{Oh}
(h^{\dagger}\,h)_{ij}={\sqrt{M_i\,M_j}\over v^2}\sum_h\,\,m_h\,
\O_{hi}^{\star}\,\O_{hj} \, ,
\ee
that can be used to recast the Eq. (\ref{eps2h}) in terms
of the $\O$ matrix elements and of $\mtt$ as
\be\label{eps2app}
\ve_2 \simeq {3\,M_2\over 16\,\pi\,\mtt\,v^2}
\left\{{\rm Im}\left[\sum_h\,m_h\,\O_{h2}^{\star}\,\O_{h3}\right]^2
+{2\over 3}\,{M_1^2\over M_2^2}\,\left(\ln{M_2\over M_1}-2 \right)\,
{\rm Im}\,\left[\sum_h\,m_h\,\O_{h2}^{\star}\,\O_{h1}\right]^2 \right\} \, .
\ee
Let us now consider three different limit cases of models
where $\rho_1\simeq 1$ and $\rho_2,\rho_3\ll 1$. The first case
is for $\rho_{2}=0$ and $\r_{22}=1$, corresponding to a small complex angle 13-rotation
\be
\O=
\left(
\begin{array}{ccc}
 \sqrt{1-\O^2_{31}}  & 0 &  -\O_{31} \\
    0 & 1 & 0 \\
  \O_{31} & 0 & \sqrt{1-\O^2_{31}}
\end{array}
\right) \, .
\ee
As we know, in this case  the $C\!P$ asymmetry $\ve_1$ is maximal. On the other hand
it is simple to see from the Eq. (\ref{eps2app}) that $\ve_2=0$ and thus the asymmetry
from the $N_2$'s vanishes and the traditional picture still holds.
The second case is for $\rho_3=0$ and $\r_{22}\simeq 1$, corresponding to
a small complex angle 12-rotation
\be
\O=
\left(
\begin{array}{ccc}
 \sqrt{1-\O^2_{21}}  &  -\O_{21}          & 0 \\
  \O_{21}            & \sqrt{1-\O^2_{21}} & 0 \\
  0 & 0 & 1
\end{array}
\right) \, .
\ee
From the Eq. (\ref{eps2app}) one can easily see that the first term vanishes and one has
\be
\ve_2\simeq {M_1\,m_{\rm atm}\over 8\,\pi\,v^2}\,\sigma\,
{M_1\over M_2}\,\left(\ln{M_2\over M_1}-2\right)\,\sin\varphi_2=
{2\over 3}\,\ve_{\rm max}(M_1)\,\sigma\,
{M_1\over M_2}\,\left(\ln{M_2\over M_1}-2\right)\,\sin\varphi_2 \, .
\ee
This result has to be compared with
$\ve_1=\ve_{\rm max}(M_1)\,\sigma\,\sin\varphi_2$
(cf. (\ref{eps1Om}), (\ref{beta})  and (\ref{hierlepphase})) showing
that $\ve_2\ll \ve_1$. Therefore, in this second case too, one has
that the asymmetry generated from the $N_2$'s can be neglected
and again the usual picture still holds.

The most interesting case is for $\rho_2=\rho_3=0$,
corresponding to a complex 23-rotation
\be\label{third}
\O=
\left(
\begin{array}{ccc}
  1  &  0   & 0   \\
  0  & \O_{22}             & \sqrt{1-\O^2_{22}} \\
  0 &  -\sqrt{1-\O^2_{22}} & \O_{22}
\end{array}
\right) \, .
\ee
The second term in the Eq. (\ref{eps2app}) vanishes while
the first, after some algebraic manipulations
\footnote{In particular we used that
$2\,{\rm Re}(\O^{\star}_{22}\sqrt{1-\O^2_{22}})\,
{\rm Im}(\O^{\star}_{22}\sqrt{1-\O^2_{22}})=-{\rm Im}(\O^2_{22})$.},
gives
 \be\label{eps2max}
\ve_2\simeq {3\,M_2\over 16\,\pi}\,{m^2_{3}-m^2_{2}\over v^2}\,{{\rm Im}(\O_{22}^2)\over\mtt} \, ,
\ee
showing that now $\ve_2$ can be large,
while on the other hand, as we know, $\ve_1=0$.
Therefore we see that,
when $\mt\ll m_{\star}$, the two possibilities of having maximum $C\!P$ asymmetry
for the $N_1$ decays or for the $N_2$ decays are complementary and
when one is maximum the other vanishes and vice versa.

The maximum value of $|\ve_2|$  in the  Eq. (\ref{eps2max}) can
be found analogously to when we maximized $\ve_1$ for
configurations with $\O_{21}=0$.
Now the role of $\O^2_{31}$ and $\O^2_{11}$
is replaced by $\O^2_{22}$ and $\O^2_{32}$ and one has to
find the maximum of $Y_{22}\equiv {\rm Im}(\O_{22}^2)$.
This can be calculated solving a set of two algebraic equations for
$Y_{22}^{\rm m}$ and for the corresponding value $X_{22}\equiv {\rm Re}(\O_{22}^2)$,
exactly equal to the Eq.'s (\ref{mtmax}) and (\ref{m3m1}) but
with the obvious replacements
$m_1\rightarrow m_2$ and $\mt\rightarrow \mtt$.
We can also introduce again a function $f(m_2,\mtt)$ related to $Y_{22}^{\rm m}$ exactly as
the function $f(m_1,\mt)$ was related to $Y_{\rm m}$ (cf. \ref{fY}), writing
\be
f(m_2,\mtt)={m_2+m_3\over \mtt}\,Y_{22}^{\rm m} \leq 1 \,
\ee
and in this way one can write for the maximum $C\!P$ asymmetry
\be\label{e2max}
\ve_2^{\rm max}=\ve_{\rm max}(M_2)\,
{m_3-m_2\over m_{\rm atm}}\,f(m_2,\mtt) \, .
\ee
Notice that this is just the bound (\ref{e1max}) for $\ve_1$,
where now $m_1$ is replaced by $m_2$ that is fixed, except for
the possibility to choose between normal and inverted hierarchy. Therefore, this time
$\ve_2^{\rm max}$ depends on whether one considers normal or
inverted hierarchy. The maximum value is obtained for
normal hierarchy because in this case $m_2\ll m_3$
and also because the function $f(m_2,\mtt)$
cuts off the asymmetry for $\mtt\leq m_2$, that means for $\mtt\leq m_{\rm sol}$
in the normal case and for $\mtt\lesssim m_{\rm atm}$ in the inverted case.
 One can understand this difference also considering that in the case of
 inverted hierarchy one has a degeneracy $m_2\simeq m_3$ and thus it is analogous
to the case of a quasi-degenerate spectrum for $\ve_1$.

The bound (\ref{e2max}) on $\ve_2$ will give rise to a lower bound  on $M_2$
analogously as the bound on $\ve_1$ determined a lower bound on $M_1$
(cf. (\ref{lbM1})). The difference is that now the function
$f(m_2,\mtt)$ is less than one, even though we are dealing with
hierarchical neutrinos and thus one has for the SM (MSSM) case
\be\label{lbM2}
M_2\gtrsim {4.2\,(2.25)\times 10^{8}\,{\rm GeV} \over f(m_2,\mtt)\,\k_{\rm f}(\mtt)} \,\, .
\ee
\begin{figure}[t]
\hspace{-40mm}
\centerline{\psfig{file=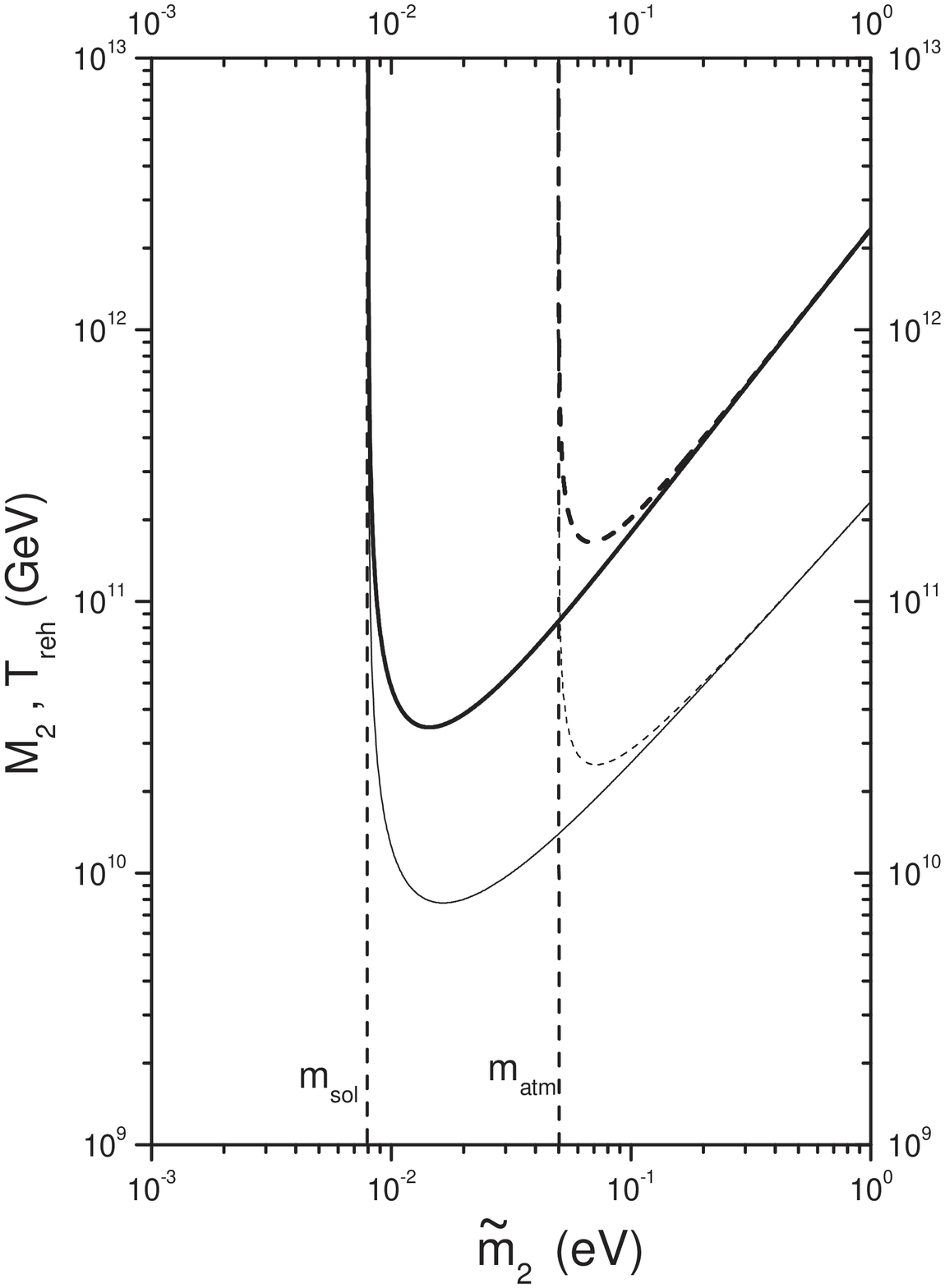,height=100mm,width=90mm}}

\vspace{-100mm}
\hspace{40mm}
\centerline{\psfig{file=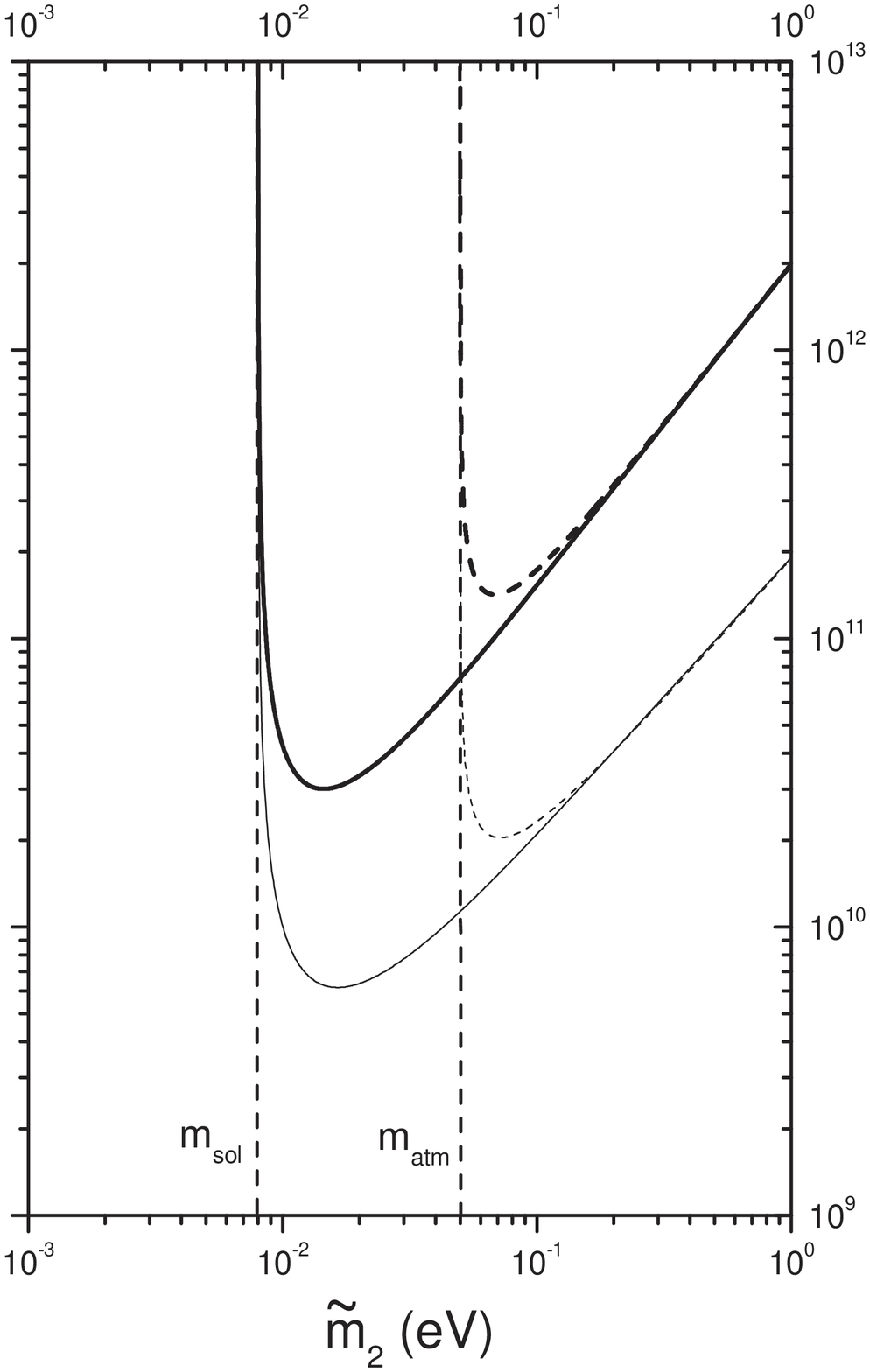,height=100mm,width=90mm}}
\vspace{-15mm}
\caption{\small lower bounds ($3\,\sigma$) on $M_2$ (solid lines) and on
$T_{\rm reh}$ (thin lines) obtained within models with $\mt\ll m_{\star}$ and
$\rho_2=\rho_3=0$. The solid lines refer to the case of normal hierarchy
while the dashed lines refer to the case of inverted hierarchy. (left) SM case;
(right) MSSM case.}
\end{figure}
This lower bound on $M_2$ is shown in Fig. 12 for both normal
(thick solid line) and inverted (thick dashed line) hierarchy.
Obviously, there will be again an associated lower bound on the
reheating temperature given by the Eq. (\ref{lbTreh}), where now
$M_1^{\rm min}$ has to be replaced by $M_2^{\rm min}$. This lower bound
on $T_{\rm reh}$ is also shown in Fig. 11 (thin lines).
The lower bounds on $M_2$ and on $T_{\rm reh}$ can be calculated again
both within the SM (Fig. 11 left) and the MSSM framework (Fig. 11 right).

Notice that within this scenario, with $\rho_2=\rho_3=0$,
there is no lower bound on the value of $M_1$, that thus can be arbitrarily
small. This occurs because we are assuming $\mt\lesssim m_{\star}$, while
 if $\mt\gtrsim 5\,m_{\star}$  the asymmetry from the $N_2$'s would be fully
 washed-out from the $N_1$ inverse decays and in this case there would not be
 any allowed value for $M_1$. Therefore, for $\rho_2=\rho_3=0$, there is a transition
 between a regime where the lower bound vanishes,
 for $\mt/m_{\star}\ll 1$, and a regime where the
 lower bound goes to infinite, for $\mt\gtrsim 5\,m_{\star}$.
In the region $m_{\star}\lesssim \mt \lesssim 5\,m_{\star}$
 there is quite a sharp transition that can be determined by a detailed
 description of the wash-out of the asymmetry produced by the $N_2$'s
 from $N_1$ inverse decays. We have not calculated the lower bound in this
 transition region, that in Fig. 8 is schematically indicated
 with a box.

Since we are assuming $\r_3,\r_2\ll \r_1 \simeq 1$, then approximately
$m_1 \simeq {v^2\,|\widetilde{h}_{11}|^2/ M_1}$. Therefore,
if $M_1$ becomes as low as the electroweak scale and because
 $m_1\ll 10^{-3}\,{\rm eV}$, then one has
 $|\widetilde{h}_{\rm 11}|\ll 10^{-7}$. This corresponds
 to a situation where the lightest right handed neutrino is decoupled
 and its interactions are switched-off, with no possibility to detect
 it in the accelerators
 \footnote{Unless one introduces some extra gauge interaction, as
 recently proposed in \cite{ky} within the context of resonant leptogenesis.
 In our case it would be even simpler, since no {\em ad hoc} motivation
 for a strong degenerate heavy neutrino spectrum is needed.}.

 Notice that one could also introduce an effective leptogenesis phase, $\sin\d_{L}^{(2)}$,
associated to the asymmetry generation from the $N_2$'s and study its suppression
when one considers cases different from the maximal production. For example
this happens if $Y_{22}\neq Y_{\rm m}$, but also if one considers
a combination of the three limit cases that we have studied, described by an $\O$
matrix that is the product of the three complex rotations.
In such a general situation one can have both a production from the $N_1$'s in the
weak wash-out regime
and a production from the $N_2$'s in the strong wash-out regime. However, not to have again
problems with the dependence on the initial conditions, one should impose that
the contribution from the $N_2$'s is the dominant one, even though not maximal.

To summarize, we have seen that the possibility of an asymmetry generated by the $N_2$'s
is realized for quite a special form of the  matrix $\O$ (cf. (\ref{third})).
Nevertheless this scenario is interesting because it solves the problems associated
with the generation of the asymmetry from the $N_1$'s with $\mt$ in the weak wash-out regime.
Indeed, we have already shown that, since $\mtt\geq m_{\rm sol}\gg m_{\star}$,
there is no dependence on the initial conditions, similarly to the typical case
of asymmetry generated by the $N_1$'s with $\mt\gtrsim 5\,m_{\star}$
\footnote{\label{footnote}
Notice that we could have also considered another possibility with
$\mt$ in the strong wash-out regime and $\mtt$ in the weak wash-out regime
such that the $N_2$ decays occur below the temperature $T_B$ where the
wash-out processes involving the $N_1$'s freeze out. In this case the wash-out would be
circumvented and the final asymmetry can also be generated by the $N_2$'s decays
or even by the $N_3$'s if $\widetilde{m}_3$ is in the weak wash-out regime
instead of $\mtt$. However, this possibility simply transfers
the problem of the dependence on the initial conditions
from the $N_1$'s to the $N_2$'s (or to the $N_3$'s).}.
  Another interesting point is that now the value of $\mt$ has not to be fine-tuned
  such that $m_1\ll \mt\ll m_{\star}$ but one has simply $\mt=m_1$. Note
  that the {\em leptogenesis conspiracy} \cite{aspects} applies also to this
  scenario. This is the observation that the
  measured atmospheric and solar neutrino mass scales lie in the correct range
   of values, between $10^{-3}\,{\rm eV}$ and $1\,{\rm eV}$, for leptogenesis
     to work. Indeed if
     $m_{\rm sol}$ and $m_{\rm atm}$ were found much larger than $1\,{\rm eV}$,
     then $\mtt\geq m_{\rm sol}$ would have yielded a too strong wash-out.
      On the other hand if
     they were found much smaller than $10^{-3}\,{\rm eV}$, then the
      maximum $C\!P$ asymmetry would have been equally smaller and
       the lower bound on $T_{\rm reh}$ at least two orders of magnitude
        more restrictive. Moreover the asymmetry production would have
         occurred in the weak wash-out regime with the related problems of
          dependence on the initial conditions.
  In conclusion, even though this scenario sits in some special
  corner of the seesaw parameter space, it represents an interesting possibility
    to be taken into account.

\section{On the hierarchy of the heavy neutrino spectrum}

Our results have been obtained under the assumption
that the spectrum of the heavy neutrinos is hierarchical and
this justified some approximations in the calculation of
the final asymmetry. We want now to discuss the conditions of validity
of these approximations.
A first aspect concerns the possibility to neglect
the asymmetry and the wash-out from the decays and inverse decays of
the two heavier neutrinos. This is possible if $M_2$ is large enough
compared to $M_1$. A conservative assumption is to impose that
the decays and inverse decays of the two lightest RH neutrinos
do not interfere with each other such that the generation
of the asymmetry from the $N_1$'s and that from the $N_2$'s proceed
independently. This holds if the generation of the asymmetry and the wash-out
from decays and inverse decays of the $N_1$'s start only after the end of the
analogous processes from the $N_2$'s.

Under this assumption and if $\mt$ lies in the strong wash-out regime,
the rate of generation of the asymmetry from the $N_1$'s, $d\eta_B/dz$,
is well approximated by a Gaussian centered
around a peak value $z_B(K_1)$ \cite{kt,bdp4}, where $K_1\equiv\mt/m_{\star}$.
More explicitly the {\em efficiency factor} $\k_{\rm f,1}$ can be written as
a Laplace integral given by
\begin{equation}
\k_{\rm f,1}  =  \int_{0}^{\infty}\,dz'\,e^{-\psi_1(z')}
\simeq  \int_{z_B(K_1)-\Delta(K_1)}^{z_B(K_1)+\Delta(K_1)}\,dz'\,e^{-\psi(z')}\, ,
\end{equation}
where $z_B(K_1)$ is that value of $z'$ where the function $\psi(z')$, determined
both by the decay rate and by the wash-out from inverse decays, has a minimum.
A second order expansion around $z_B(K_1)$ makes possible to approximate the
exponential with a Gaussian centered at $z_{B}(K_1)$.
This procedure is valid only if the inverse decays from the $N_2$'s
are inefficient for $z\gtrsim z_B(K_1)-\Delta(K_1)$, otherwise there would
be an additional contribution to $\k_{\rm f,1}$ coming from the
$N_2$ inverse decays wash-out.
If $\mtt$ lies in the strong wash-out regime too and if one neglects
the wash-out from $N_1$ inverse decays
\footnote{This is a conservative assumption, an account of
this effect would go into the direction of relaxing the final condition
on the ratio $M_2/M_1$.},
then the generation of the asymmetry
from the $N_2$'s will be also described by a Gaussian centered around
a value $z\simeq z_B(K_2)\,M_1/M_2$, with $K_2\equiv \mtt/m_{\star}$. This means that
for values of $z\simeq [z_B(K_2)+\Delta(K_2)]\,M_1/M_2$ the generation of the
asymmetry from the $N_2$ decays and the wash-out from the $N_2$ inverse decays
can be neglected. Therefore, they will not influence the final
value of the asymmetry, if the peaks of the two Gaussian are sufficiently separated,
that means if $[z_B(K_2)+\Delta(K_2)]\,M_1/M_2\lesssim z_B(K_1)-\Delta(K_1)$.

Let us impose a separation of the two peaks of about $3\,\sigma$
that guarantees  a precision much less than $10\%$ in the calculation
of the asymmetry when $N_2$ decays and inverse decays are neglected. Using
the expression (\ref{zB}) for $z_B(K)$ and considering that the width of the
Gaussian is approximately constant with $K$ and given by
$\sigma\simeq 1.5$ \cite{bdp4}, this implies to have
$\Delta(K_1)\simeq \Delta(K_2)\simeq 1.5\,\sigma \simeq 2$. Therefore,
we arrive to the following very conservative condition
\be
{M_2\over M_1}\geq {z_B(K_2)+2\over z_B(K_1)-2} \, .
\ee
 Taking the most conservative choice of values
\footnote{This choice corresponds to have $N_1$ decays in the strong wash-out
regime and $\mtt\sim 10\,{\rm eV}$. This large $\mtt$ value can be
obtained if $\mtt \sim m_{\rm atm}\,|\O_{32}^2|$ and  $|\O_{32}^2|\sim 10^3$.
Since $M_3\gtrsim 10^{10}\,{\rm GeV}$,
from the eq. (\ref{Omegah}) one can check that values $\mtt\gg 10\,{\rm eV}$
would imply Yukawa couplings much larger than 1.}, $K_1\simeq 10$ and $K_2\simeq 10^4$,
the RH side of this inequality is maximized and one finds that:
{\em if $M_2/M_1\gtrsim 5$, then the effect of $N_2$ decays
and inverse decays on the final asymmetry can be safely neglected}.
If $\mt$ is in the weak wash-out regime, then one can have a situation
as described in Section 6, while if $\mtt$ is in the weak wash-out one can have
a situation as described in footnote \ref{footnote}.

The second aspect to be considered is the $C\!P$ asymmetry
and in particular the conditions of validity of the approximations
that from Eq. (\ref{eps1g}) lead to Eq. (\ref{eps1h}).
The first one can be written as
\be\label{eps1gbis}
\varepsilon_1 \simeq  {3\over 16\,\pi}
\sum_{i=2,3}\,{{\rm Im}\,\left[(h^{\dagger}\,h)^2_{1i}\right]\over
(h^{\dagger}\,h)_{11}} \,{\xi(x_i) \over \sqrt{x_i}} \, .
\ee
where $x_i\equiv M^2_i/M_1^2$ and the function $\xi(x)$, given by \cite{bdp2}
\be
\xi(x)={2\over 3}\,x\,
\left[(1+x)\,\ln\left({1+x\over x}\right)-{2-x\over 1-x}\right] \, ,
\ee
approaches $1$ for $x\gg 1$.
After some easy algebra, the Eq. (\ref{eps1gbis}) can be re-casted as
\be\label{eps1gref}
\ve_1=\xi(x_2)\,\ve_1(M_1,m_1,\mt,\O_{j1}^2)+[\xi(x_3)-\xi(x_2)]\,\Delta\ve_1 \, ,
\ee
showing that in the limit $x_2,x_3\rightarrow \infty$
one recovers the result $\ve_1\simeq \ve_1(M_1,m_1,\mt,\O_{j1}^2)$
(cf. \ref{eps1h}), while, for finite $x_2,x_3$, there is both an enhancement
$\xi(x_2)$ of the usual term, plus an additional contribution
$[\xi(x_3)-\xi(x_2)]\,\Delta\ve_1$, where
\be\label{Deps1}
\Delta\ve_1\equiv {3\over 16\,\pi}\,{{\rm Im} [(h^{\dagger}\,h)^2_{13}]\over
(h^{\dagger}h)_{11}}
\,{1\over \sqrt{x_3}} = \ve_{\rm max}(M_1)\,
{{\rm Im}[\sum_h\,m_h\,\O_{h1}^{\star}\O_{h3}]^2\over m_{\rm atm}\,\mt}  \, .
\ee
In the second expression we have used the relation
that connects $h$ to $\O$ (cf. (\ref{Oh}))
and the definition of $\ve_{\rm max}$ (cf. (\ref{e1maxM1})).
In this way we can write
\be\label{xieps}
\xi_{\ve}\equiv {\ve_1 \over \ve_{\rm max}(M_1)}
=\xi(x_2)\,\b_{\rm max}(m_1,\mt)\,\sin\d_L+[\xi(x_3)-\xi(x_2)]\,
{{\rm Im}[\sum_h\,m_h\,\O_{h1}^{\star}\O_{h3}]^2\over m_{\rm atm}\,\mt} \, .
\ee
First of all let us notice that in this general case the asymmetry
depends on 4 additional parameters, the masses $M_2$ and $M_3$
and the two $\O$ parameters that were disappearing within the
assumption of heavy  hierarchical spectrum.
If we use the  parametrization Eq. (\ref{O}) for $\O$,
then these two additional parameters can be identified with the
real and the imaginary part of $\O_{22}$.

It is instructive to study $\xi_{\ve}$ in two  limit cases.
For $x_2=x_3$ (i.e. $M_2=M_3$) one has $\Delta\ve_1=0$
and the enhancement of the asymmetry, compared to the hierarchical case,
is simply described in terms of $\xi(x_2)$ \cite{bdp2}. Therefore we see
that in this case the final asymmetry will depend just on one
additional (seventh) parameter compared to the case when a full
hierarchical heavy spectrum is assumed.
  Like for the effect of asymmetry and wash-out from
the $N_2$'s, one has again that if $M_2\gtrsim 5\,M_1$,
then the precision of the approximation
of heavy hierarchical spectrum is much below
\footnote{For $M_2=5\,M_1$, one has $\xi(x_2)-1\simeq 0.02$.}
$10\%$.

The second limit case is obtained for $x_3\rightarrow\infty$
such that the difference $\xi(x_3)-\xi(x_2)\simeq 1-\xi(x_2)$ is
maximum with respect to $x_3$.
 It is useful to  calculate $\D\ve_1$
 in the case of maximal effective phase and full hierarchical
neutrinos such that $\O_{21}=X_3=0$;
the general expression (\ref{xieps}) becomes simply
\be
\xi_{\ve}=\xi(x_2)+[\xi(x_2)-1]\,[X_{22}+Y_3\,Y_{22}] \, .
\ee
One can see that a crucial role is clearly played by the imaginary and by the
real part of $\O_{22}$.
If $X_{22}+Y_3\,Y_{22}=-1$, one has the curious result
that $\ve_1=\ve_{\rm max}(M_1)$, irrespective of the value of $x_2$.
Thus we see that the enhancement of the $C\!P$ asymmetry, when
$M_2\rightarrow M_1$, is not unavoidable,  but a possibility.
In general, if $X_{22}+Y_3\,Y_{22} \sim {\cal O}(1)$ and if $M_2\gtrsim 2\,M_1$,
one has that the first term in the (\ref{eps1gbis}) dominates, while
if $X_{22}+Y_2\,Y_{22}$ is much larger than one, then a dominance of the second term
is possible.
However, if one imposes $M_2\gtrsim 5\,M_1$, large values $\xi_{\ve}\gtrsim 2$,
necessary to evade significantly the lower bounds on $M_1$ and on $T_{\rm reh}$,
rely on unnaturally huge value of $X_{22}+Y_3\,Y_{22}$, implying huge values
of $\mt$ or of $\mtt$ or both.
Analogous results have been obtained also in \cite{dk}.


Notice that in  general one can say that the two particular cases we have studied,
$x_2=x_3$ and $x_3=\infty$,
can be regarded as limit cases of two different classes of
RH neutrino spectrum:  {\em inverted heavy neutrino spectra},
with $M_3^2-M_2^2<M_2^2-M_1^2$,
and {\em normal heavy neutrino spectra}, with $M_3^2-M_2^2>M_2^2-M_1^2$.
For small imaginary parts and small values of $\D\ve_1$ the second term
in the Eq. (\ref{xieps}) is small and $\xi_{\ve}$ is the
same in both cases, while for large imaginary parts the case of normal
heavy neutrino spectra gives rise to a larger enhancement arising from $\D\ve_1$.
However as far as $M_2\gtrsim 5\,M_1$ the enhancement can be relevant in changing
considerably the lower bounds on $M_1$ and on $T_{\rm reh}$ only for large values
of the imaginary parts.

Another issue concerns the {\em effective leptogenesis phase}.
The second term in the Eq. (\ref{xieps}) is not maximal for the same
configurations that maximize the first one and thus the effective leptogenesis phase
$\sin\d_L$ is related only to the first term.
In particular
we have seen that the second term is sensitive also to the values of $X_{22}$ and $Y_{22}$.
Here we are not interested in a systematic analysis of $\D\ve_1$,
but it is interesting to describe two particular cases.
 First of all notice that if $\mt=m_1$, that is equivalent
to assume an $\O$ matrix as in the Eq. (\ref{O23}), then not only
the first term vanishes, because $f(m_1,\mt)=0$,
but one can check that $\D\ve_1=0$ too. This is important
because,  otherwise the results that are usually presented in the limit
$\mt\rightarrow m_1$, would have been valid only under special conditions.
On the other hand we can consider a simple case showing that
when $\sin\d_L=0$ the second term does not necessarily vanish.
Let us take $X_2=Y_2=Y_3=0$, such that $\sin\d_L=0$, and
let us also assume for simplicity the case of full hierarchical neutrinos,
such that $\mt=m_3\,X_3$. In this case one can easily arrive to the expression
\be
\xi_{\ve}=[\xi(x_3)-\xi(x_2)]\,(1-X_3)\,Y_{22} \, ,
\ee
showing that the second term does not vanish in general when
$\sin\d_L=0$. However, once more, we can see that if $M_2\gtrsim 5\,M_1$,
then one has to require $Y_{22}\gg 1$ to have $\xi_{\ve}\gtrsim 1$.

Summarizing, we can say that if the condition $M_2\gtrsim 5\,M_1$ holds,
then the usual approximations work very well with a great precision,
except for some situations involving large imaginary parts of the $\O_{ij}$'s.

If one considers quasi-degenerate light neutrinos with $m_1\gtrsim m_{\rm atm}$,
then, while the first term in the (\ref{xieps}) gets suppressed as
$\propto (m_3-m_1)/m_{\rm atm}$,
the second does not and therefore it can become dominant more easily, especially if one
considers a `normal' heavy neutrino spectrum with $x_3\gg 1$  and finite $x_2$ \cite{hambye}.
  This is relevant in connection with the neutrino
mass upper bound. One has to say however that the possibility
of having $m_1\gtrsim m_{\rm atm}$ holds only for a restrictive
choice of the parameters. In particular for reheating temperature
$T_{\rm reh}\ll 10^{11}\,{\rm GeV}$ the bound falls in the hierarchical
regime and thus one does not expect important corrections, but further
investigation is needed on this point.

\section{Summary and final discussion}

Leptogenesis is not only an attractive way to explain the
baryon asymmetry of the Universe but also a powerful
cosmological tool to get a unique information on the seesaw parameters.
It is intriguing that the $C\!P$ asymmetry underlying the origin
of the matter-anti matter asymmetry of the Universe could be related to a
sort of `dark side' of the seesaw parameter space, the orthogonal
matrix $\Omega$, that escapes Earth experiments but leaves a
cosmological imprint.

Our investigation could cast light into this dark side  thanks to
a geometrical representation of the orthogonal
seesaw matrix, resembling the use of unitarity triangles in the
study of $C\!P$ violation in the weak interactions of quarks.
Interesting features of leptogenesis and of its implications for neutrino
mass models have emerged.

With the use of the seesaw geometry it has been possible
to calculate the $C\!P$ asymmetry  bound that gives rise to
the leptogenesis constraints on the neutrino masses and on the
reheating temperature, fully characterizing the models that saturate the bound.
At the same time it has been possible to study the phase
suppression in general models compared to the  case of maximal asymmetry
and in this way we could place an interesting lower bound (cf. (\ref{lbsind})).

The results have specified which neutrino mass models can access
different range of values of $M_1$ and $T_{\rm reh}$. In particular,
\begin{itemize}
\item the lowest allowed range for
$M_1,T_{\rm reh} \sim (10^{9}-10^{10})\,{\rm GeV}$ is actually accessible only
to a particular class of neutrino mass models where the lightest neutrino
mass is dominated by the contribution of the lightest RH neutrino, a
class of models that cannot be easily
motivated within models where neutrino masses
resemble quark mass models \cite{kingrep}
\footnote{This possibility can be however realized
in more sophisticated models of fermion masses, as for example
in a recent 6 dimensional orbifold $SO(10)$ model \cite{asakacovi}.};
\item models with the lightest RH neutrino $N_1$ dominantly contributing
to one of the two heavier (light) neutrino masses, $m_2$ and $m_3$,
that are typically considered to explain large
mixing angles resembling quark mass models, undergo both phase suppression and
strong wash-out such that the lower bounds become  more restrictive:
$M_1\gtrsim 1.5\times 10^{11}\,{\rm GeV}$ (within SM),
$T_{\rm reh}\gtrsim 2\times 10^{10}\,{\rm GeV}$ (within MSSM) if
$N_1$ dominantly contributes to $m_3$ and $M_1\gtrsim 4\times 10^{10}\,{\rm GeV}$,
$T_{\rm reh}\gtrsim 7\times 10^{9}\,{\rm GeV}$ (within MSSM) if
$N_1$ dominantly contributes to $m_2$.
\end{itemize}
We have also studied whether these restrictions can be circumvented
when the usual assumption of hierarchical heavy neutrino
spectrum is dropped. A significant enhancement of the final asymmetry
relies on the possibility either of a  very fine tuned degeneracy such that
$M_2\simeq M_1$, either on large values of the imaginary parts of the $\O$
parameters, we have seen the special role played in particular by $\O_{22}$,
or on some combination of the two things. A better understanding of the relevance of
these models is certainly needed, however at the moment
it seems that their feasibility relies on some {\em ad hoc} model engineering.

In conclusion the results we have obtained can be regarded
either as something that exacerbates
the problems of thermal leptogenesis related to large values of $M_1$ and of $T_{\rm reh}$,
the most popular being the gravitino overproduction within
traditional supergravity models, either as a support of thermal
leptogenesis to less typically considered possibilities for neutrino models or,
a third possibility, as a support to inflationary models with quite large
reheating temperatures.

Further  investigations are needed to understand which one of these possible
interpretations is the correct one.
If it will prove possible to built realistic models
with the lightest neutrino mass dominated by the lightest RH neutrino, then our
analysis has revealed also another interesting possibility,
namely that the  final baryon asymmetry could be actually produced not from
the decays of the lightest RH neutrinos, the $N_1$'s, as usually considered,
but from those of the second lightest, the $N_2$'s. This is a particular
scenario working for $\mt\lesssim m_{\star}$ (cf. (\ref{third})), that however
does not suffer of the problem of the dependence on the initial conditions.

In conclusion, our analysis confirms that thermal leptogenesis is a
very predictive model of baryogenesis, implying tight and non trivial
correlations between the final baryon asymmetry and neutrino mass models. It is impressive
the amount of implications that is possible to derive by requiring the explanation
of just one single observable, the baryon asymmetry,
within an eighteen parameter model like the seesaw.
 A large variety of neutrino mass models
can explain the baryon asymmetry within the seesaw mechanism but non trivial
constraints apply and the value of the reheating temperature emerges as one of the most
crucial parameters on which unfortunately we have little experimental information.
Future investigations should be hopefully able to understand whether these constraints
will somehow point to a particular neutrino mass model, able to explain
all observations simultaneously, from neutrino experiments to the baryon asymmetry, or
some severe conflict will arise whose solution will require one of the
many possible extensions of the minimal leptogenesis model with the problem
to discriminate in favor of one of them.
At the moment it is quite  remarkable how the oldest minimal leptogenesis model
is still alive, having successfully  passed important
tests represented by the values of the neutrino mixing scales, that well allows to talk of
a {\em leptogenesis conspiracy} \cite{aspects} and with the future perspective that
this conspiracy can become even tighter if the absolute neutrino mass scale upper
bound,  $m_i < 0.1\,{\rm eV}$, will be confirmed by the experiments.

\vspace{2mm}
\noindent
\textbf{Acknowledgments}\\
It is a pleasure to thank W.~Buchm\"{u}ller, M.~Pl\"{u}macher and G.~Raffelt
for useful comments and discussions. I wish to thank  M.~Pl\"{u}macher
also for a careful reading of the manuscript and K.~Turzynski
for comments and for having pointed out ref. \cite{ct}.
\noindent

\section*{Appendix}

Let us show that starting from a generic configuration
it is always possible to find, for a fixed value of $\mt$, a configuration with
higher value of $\beta$ and $Y_2=0$. There are two cases to be distinguished.

The first case occurs starting from a configuration with $Y_2<0$ (i.e. $Y<Y_3$).
An example of such a configuration is shown in Fig. 3 with dotted arrows
\footnote{Notice that one can always assume $|\varphi_2|\leq |\varphi_3|$, since
this has always higher value of $\b$ compared to a dual configuration
with $|\varphi_2|\leq |\varphi_3|$ obtained from a rigid rotation
of $\O_{21}^2+\O_{11}^2$ around the axis that passes through the
points $(X_3,Y_3)$ and $(1,0)$.}.
It is then simple to construct a configuration with $Y_2=0$ ($Y=Y_3$),
same value of $\mt$ and higher value of $\beta$
(or equivalently of the $C\!P$ asymmetry).
Let us fix $\Omega^2_{31}$, that  means the values
of $X_3$ and $Y_3$, and let us pass to a configuration with a smaller $\rho_2$.
In this case one has that the sum
$\rho_2+\rho_1=A_3/m_1-\rho_2\,(m_2/m_1-1)$, with constant $A_3$, where
$A_i\equiv \mt-\rho_i\,m_i$, increases. This
implies that decreasing $\rho_2$ with a continuous transformation,
one necessarily arrives to a configuration with $Y_2=0$.

The second case occurs starting from a configuration with $Y_2>0$.
An example is shown in Fig. 3 with dashed line arrows
\footnote{This time, analogously to the first case, one can always assume
$\varphi_3\geq\varphi_2$.}.
Let us fix this time $\O^2_{11}$, that means the values of $X$ and $Y$.
In this case one can again pass to a configuration with smaller
$\rho_2$, same value of $\mt$ and higher value of $\beta$.
However in this case one can end up either again with a configuration with $Y_2=0$,
shown in Fig. 3 with the solid line arrows, or with a configuration such that
$\Omega_{31}^2$ and $\Omega_{21}^2$ are aligned,
corresponding to $\varphi_3=\varphi_2$ and
shown in Fig. 4 with dotted arrows. This because if one decreases $\rho_2$ keeping
$\mt$ constant, then the sum $\rho_2+\rho_3=A_1/m_3+\rho_2\,(1-m_2/m_3)$ necessarily
decreases while, as we will show in a moment, $Y_3$ increases and this implies
that $\beta$ increases too. Let us show that $Y_3$ increases for
decreasing $\rho_2$. It is easy to find an expression of $Y_3$
as a function of $\rho_2$ using simple geometrical relations, obtaining
\be
Y_3={Y(X^2+Y^2+\rho\,\Delta\rho)\pm X\,
\sqrt{[\rho^2-(X^2+Y^2)]\,[X^2+Y^2-\Delta\rho^2]}\over 2(X^2+Y^2)} \, ,
\ee
where we defined $\rho\equiv\rho_2+\rho_3$ and $\Delta\rho=\rho_3-\rho_2$.
If $Y_2\leq 0$, as we are assuming, it is then easy to show that
$\partial Y_3/\partial\rho_2\leq -2\,Y(\rho-\Delta\rho)\leq 0$,
showing that $Y_3$ increases when $\rho_2$ decreases.

If the initial configuration is such that $A_1/m_3 \geq X^2+Y^2$,
then it is possible to decrease $Y_2$ until this vanishes,
otherwise one ends up with a configuration such that $\varphi_2=\varphi_3$.
 In this second case one can however perform a different kind of
 transformation fixing this time
the values of $X$ and $X_3$ (other than that of $\mt$) and if one
decreases $Y_2$ then $\beta$ increases.
 This can be shown easily taking the derivative of the Eq. (cf. (\ref{mt1XY}))
 with respect to $Y_2$ and finding an expression for $dY_3/dY_2$ given by
\be
{dY_3\over dY_2}= -{\ds m_2\sin\varphi_2+m_1\sin\varphi_1
\over\ds m_3\sin\varphi_3+m_1\sin\varphi_1}
\leq -{m_2\over m_3}\, .
\ee
From the expression (\ref{bY}) we can then easily find that
\be\label{negative}
m_{\rm atm}\,\mt\,{d\beta\over dY_2}=\Delta m^2_{21}-\Delta m^2_{31}\,{dY_3\over dY_2} < 0 \, .
\ee
Thus the $C\!P$ asymmetry increases when $Y_2$ decreases. One can try to decrease $Y_2$
until it vanishes. In this case one can again apply the previous result valid for
configurations with $Y_2=0$. However this is not guaranteed to be possible, since
the result (\ref{negative}) is valid only for an infinitesimal transformation starting
from $\varphi_2=\varphi_3$ and it is not in general possible to decrease
$Y_2$ of a finite arbitrary quantity, while the $C\!P$ asymmetry increases and
$\mt$ is kept constant. However in any case such a transformation
brings again to a configuration with $\varphi_3>\varphi_2$.
One  can then again shorten $\rho_2$ through a transformation with fixed values of $Y$ and $X$ and
again this can bring either to a configuration with $Y_2=0$ or to one with
$\varphi_3=\varphi_2$. In the second case the procedure can be repeated recursively until
a configuration with $Y_2=0$ is obtained anyway. In Fig. 4 we give a very schematic example
of such a procedure. We have thus demonstrated that, starting from any generic
configuration, there is always another configuration with the same value of $\mt$, higher
value of $\beta$ (or equivalently higher value of the $C\!P$ asymmetry) and $Y_2=0$.

\end{document}